**Sea Spider development: How the encysting *Anoplodactylus eroticus* matures from a bouyant nymph to a grounded adult**

Amy Maxmen

In order to understand how animals evolved over time, biologists must learn how their body parts form during their development. The following is a detailed description of how one species of sea spider transforms from a hatchling to an 8-legged adult. It is a chapter from my doctoral thesis on the evolution and development of sea spiders (pycnogonids), and arthropods in general, completed in 2005 in the Organismic and Evolutionary Biology department of Harvard University.

**Why study sea spiders?**

Pycnogonids, or sea spiders, comprise a primitive lineage of arthropods. As such, they hold potential to reveal insights into arthropod evolution. The phylogenetic position of pycnogonids among the four major arthropod classes (chelicerates, myriapods, crustaceans, and insects) remains ambiguous. Recent phylogenetic analyses have supported their position as either basal chelicerates or as a separate, fifth major lineage of extant arthropods. A central difficulty in placing the pycnogonids is that they exhibit a combination of characters serving to variously ally and distance them to other arthropods (as well as non-arthropod ecdysozoans). A second problem is that the group is understudied, and therefore pycnogonid characters added to cladistic data matrices are often absent, incomplete, or mistaken.

Arthropod phylogeny and morphological evolution can only be understood through comparisons in which homologies are adequately assessed (i.e. signal overcomes noise). For this reason, the homology concept reoccurs at every chapter in comparative discussions of observed pycnogonid characters. I devote particular attention to the front claws (chelifores) since the classification of pycnogonids with chelicerates relies on the presumed homology of chelifores and the front appendages of chelicerates. Conversely, I've highlighted the lack of an obvious homolog to the arthropod labrum in pycnogonids, because without an equivalent structure, there is increased support for a separate lineage of pycnogonids which split off early from the ancestor of all other extant arthropods. Furthermore, unlike the typical arthropod germ band in which the embryonic body is initially patterned, *A. eroticus* is patterned in two separate stages. Larvae hatch with three appendage-bearing segments corresponding to the adult head, and the post-cephalic region is patterned in sync at a later stage.

Understanding the dynamics of pycnogonid body patterning is a prerequisite for bringing pycnogonids into contemporary studies in arthropod evolution and development ('evo-devo') that generally compare early stages at a molecular level.

**Post-embryonic development of the sea spider *Anoplodactylus eroticus* (Arthropoda, Pycnogonida)**

**Abstract**

Disagreements concerning pycnogonid relations to other arthropods, including homologous characters, is partly a product of too few primary observations of pycnogonid anatomy and development. This investigation of post-embryonic development of the pycnogonid *Anoplodactylus eroticus* employs multiple techniques of anatomical observation in order to thoroughly document the life cycle. Morphogenesis is described as a series of stages identified by examination of live, freshly collected specimens under brightfield microscopy with Nomarski optics and of fixed specimens with the use of scanning electron microscopy and fluorescence microscopy to detect cross-reactive molecular markers. After the second post-embryonic stage, larvae of *A. eroticus* burrow within a hydroid and undergo morphogenesis. Larvae emerge from the hydroid and simultaneously molt into the juvenile stage. Over the course of post-embryonic development there are eight stages preceding the mature adult. All structures, except for the anteriormost appendages, the chelifores, undergo some degree of transformation. Chelifores are present prior to hatching and remain mobile over the course of development. Some larger issues important to arthropod evolution are addressed, such as the equivalent of a germband and labrum in pycnogonids. Post-embryonic development of *A. eroticus* provides an example counteracting previous reports of anamorphic development and a four-segmented head in the pycnogonid ground pattern, findings that were extrapolated to fit the ground pattern of Arthropoda.

**Introduction**

Sea spiders comprise the arthropod taxon Pycnogonida (latin for "knobby-knees"), a group named for their characteristic four to six pairs of spindly legs. Pycnogonids are exclusively marine and exhibit an unusual array of ecological, behavioral, and morphological characters such as male parental care, a prominent sucking proboscis with a terminal Y-shaped mouth, a single-segmented abdomen (anal tubercle), and a pair of ventral appendages (ovigers) on the cephalic segment used to carry the young. Due to the number of autapomorphies, the phylogenetic placement of pycnogonids has wobbled[1,2] leading to a history of uncertain placements with the crustaceans[3], the chelicerates[4,5], or, alternatively, as a separate group of arthropods[6-11]. Morphological similarities between the aquatic larvae of the pycnogonid protonymphon and the crustacean nauplius were used to suggest affinities between these groups[3,12]. However, unlike the crustacean nauplius, protonymphon appendages are always uniramous, and as adults share little else to unite them with crustaceans. Resemblance of the anteriormost chelate pair of pycnogonid appendages (chelifores) to those in horseshoe crabs and spiders (chelicerae) has been used to suggest a close relationship between pycnogonids and chelicerates[13-16]. However, the homology of these anterior appendages has been recently challenged[17]. Unlike chelicerates, pycnogonids do not have a body composed of a fused cephalothorax (prosoma) and abdomen (opisthosoma).

Recent phylogenetic analyses employing molecular data suggest that pycnogonids are either basal chelicerates [18-22], form a distinct lineage in a 'chelicerate + myriapod' clade[22], or are sister taxon to the remaining arthropods [23-25]. The fossil record indicates that pycnogonids branched off early during the emergence of stem-group arthropods with earliest larvae identified from the Upper Cambrian [26] (about 490 Myr ago) and a crown-group (or near-crown group) adult from the Siluran period[13] (about 425 Myr ago). The ongoing controversy over the placement of pycnogonids is in large part due to the lack of morphological and developmental information about the group.

Various stages of the 'encysting' pycnogonid examined here, *Anoplodactylus eroticus*, have been collected from within and upon the hydroid, *Pennaria disticha* (Hydroida: Halocordylidae), extensively for two years in an effort to understand postembryonic development as thoroughly as possible. Pycnogonid postembryonic development has been categorized into four general types[5] based on life history differences termed 'encysting', 'free-swimming' (or 'typical'), 'attaching', and 'atypical'. In 'encysting' development, the first stage protonymphon hatches from the egg and leads a free-swimming existence prior to shedding its two pairs of post-cheliforal larval appendages, and burrowing into a cnidarian host where it will undergo a series of transformations, emerging as a juvenile. Depending on the type of cnidarian host, the protonymphon may or may not form a cyst inside the host[5,27]. Larvae of one species, *Anoplodactylus petiolatus,* have been found encysting within the manubrium of the *Obelia* medusae[28]. In contrast to encysting developers which form three post-cephalic segments, each with a pair of walking limb primordia, simultaneously, the body segments and limbs of free-swimming developers are added sequentially[29-31]. In 'attaching' postembryonic development, the Stage I protonymphon does not hatch from the egg, but emerges at a later stage and remains attached to the male ovigers. Like the free swimming protonymphon, leg primordia are added one at a time with each molt (represented by *Propallene longiceps*[32] and *Austropallene conginera*[33]). In 'atypical' development, the protonymphon is a parasite on another non-cnidarian host, such as a clam, nudibranch, or polychaete[5] and limbs are modified to suit the parasitic lifestyle[34]. Similar to encysting species, the adult appendages grow in sync[34].

The encysting lifecycle has been recognized for over a century. Protonymphae typical of *Phoxichilidium* and *Anoplodactylus*, with "long tendril-like extensions" were first noticed by Gegenbauer in 1854 among hydroids, and in 1862, Hodge showed that these protonymphae would encyst within the hydroid for the duration of post-embryonic transformation into the juvenile[35]. Despite a century of reports, the complete lifecycle of encysting pycnogonids has never been directly observed due to complications in maintaining healthy host cnidarians in the laboratory. Therefore descriptions are based on isolated stages that have been variously collected from the field[12,27,29,36-38], rendering descriptions of development rare and often incomplete.

Understanding encysting development is important in order to make comparisons between other arthropods and pycnogonids based on observations drawn from diverse pycnogonids. Vilpoux and Waloszek[30] set a high standard of description based on scanning electron micrographs (SEM) of the free-swimming developer, *Pynogonum litorale*, Their style of description has been employed here with the encyting development of *A. eroticus* from protonymphon to adult. Additional modern tools and techniques were used for recording observations, including immunohistochemistry in conjunction with confocal microscopy (cLSM) and brightfield microscopy with Nomarski optics (differential interference contrast (DIC)). This investigation does not include a phylogenetic analysis, and therefore the position of pycnogonids among arthropods is excluded from discussion. However, characters applicable to future phylogenetic analyses, such as the pycnogonid head, labrum, appendages, and mode of development are critically examined.

**Methods and Materials**

In vivo light microscopy and SEM of fixed specimens was used to document all stages. Visualization of nuclei in fixed Stage I and Stage V larvae was made possible by DAPI or Propidium Iodide staining. Musculature was selectively stained by labeling F-actin with flourescently conjugated phalloidin. We also used two cross-reactive antibodies as molecular markers. Specifically, anti-acetylated tubulin was used to mark neural structures in the protonymphon and juvenile, and the product of the gene *Distal-less* (*Dll*), known for possessing a conserved expression pattern in the distal section of developing arthropod appendages[39-42] and also being associated with sensory organ development[43], was observed in the protonymphon.

*Collection and preservation*

*Anoplodactylus eroticus* was collected on "Christmas tree" hydroids, *Pennaria disticha*, found at a depth of 1-5 m on suspended objects such as anchor lines and boat bottoms at Kewalo Basin Harbor, Honolulu, Hawaii, throughout the year during 2004-2006. Specimens were sorted in the laboratory. Encysting larvae were located by identifying swollen hydranths and removing larvae with insect pins and forceps under a dissecting microscope. Protonymphae were collected primarily by allowing embryos removed from brooding males to rest in 22 μm Millipore filtered sea water in 35-mm Petri dishes at room temperature (27ºC). After 1-3 days, approximately 25-50 protonymphae would hatch per male brood. Live specimens were observed under a Zeiss Axio Imager.Z1. Specimens to be used for SEM were fixed in either 3.7% formaldehyde in phosphate-buffered saline solution (1 x PBS) or Boiun's fluid (300μl Picric Acid (saturated), 100μl 37% formaldehyde, 20μl glacial acetic acid). Bouin's fluid made a positive difference in post-embryonic stages in which the cuticle is not as robust as in the adult. After fixation, the material was dehydrated in an ethanol series (75%, 85%, 95%, 97%, 100%). For DAPI, Propidium Iodide, Phalloidin, and antibody staining, specimens were fixed in 3.7% formaldehyde in 1 x PBS for <10 minutes at room temperature while rocking gently, then washed into PBS with 0.1% Triton X-100 (PBT) three times for 5 min, and 1 x PBS once for 5 min.

*SEM*

Fixed specimens were stored in 100% ethanol at 4°C. Prior to observation, specimens were critical-point-dried, mounted, and gold-coated, following standard procedures. Observations were made on a Quanta 200 ESEM at the Harvard University Herbaria.

*Nuclear Staining*

For visualization of nuclei by flourescence microscopy, fixed specimens were incubated in 50% glycerol/1 x PBS containing 1mg/ml DAPI overnight at 4°C. Propidium Iodide was alternatively used (based on the availability of cLSM) to label cell nuclei. Fixed specimens were incubated in Propidium Iodide and RNAse solution in PBT at room temperature for one hour. Specimens were then washed 3 times for 15 min with PBT at room temperature while rocking.

*Musculature staining*

Fixed specimens were washed into PBS, and incubated in FITC-labeled phalloidin (BODIPY FL phallacidin, Molecular Probes #B-607; 2μl 6.6 μM stock solution into 300μl PBS) in the dark for 2h at room temperature and subsequently washed 3 times for 5 min in PBS.

*Immunohistochemistry*

Antibody detection of anti-acetylated tubulin with an HRP color reaction was performed on fixed protonymphae as described previously in Patel (1998)[44] and detection of Dll and Elav on fixed protonymphae and anti-acetylated tubulin on fixed juveniles with fluorescent secondary antibodies as described in Dickinson[45] with the following modifications. After fixation in 3.7% formaldehyde, protonymphae were bath-sonicated for 2, 4 sec pulses and juveniles sonicated for 4, 4 sec pulses at a low setting. Staining of neuronal structures was obtained with the monoclonal anti-acetylated tubulin antibody, and independently, with Elav (for each, mouse IgG; Developmental Studies Hybridoma Bank) at dilutions of 1:5 and 1:200, respectively, in block solution (PBS containing 0.1% Triton X-100 and 5% normal goat serum) overnight at 4°C. Antibody detection of Dll was obtained with the polyclonal anti-Dll antibody (rabbit IgG; Panganiban *et.al* 1995[46]) at a dilution of 1:100 overnight at 4°C. Specimens labeled with the Elav and Dll antibody were incubated for 5h at room temperature (27°C) in secondary antibodies, Cy3-conjugated goat anti-mouse (Jackson ImmunoResearch Laboratories) and Alexa-488-conjugated donkey anti-rabbit (Molecular Probes) were applied for 5h at room temperature, at a dilution of 1:250. Omission of the primary antibody abolished staining.

### *Images*

Images of DAPI and anti-tubulin immunostaining were captured on a Zeiss Axio Imager.Z1 at 3-9 focal planes. Images of propidium iodide, phalloidin, and the flourescently-conjugated antibodies were taken with a Zeiss LSM510 META confocal laser-scanning microscope.

### Results

Post-embryonic development of *Anoplodactylus eroticus* could be divided into nine stages, the first being the protonymphon and the ninth being the mature adult (**Table 1**).  Stages are defined by limb development, since total body length is not incremental and varies within stages. The number of instars, each separated by a molt, could not be determined by this work as we were unable to monitor individual specimens over significant periods of time due to their parasitic life history. Cuticular condition has been used as a proxy for the stage in the molting cycle, however, since the cuticle is also affected by the fixatives used to prepare the specimen, wrinkling may be an artifact of preservation and therefore should not be used as a sole basis for inference.

**Table 1.** Body measurements and major features characterizing developmental stages of *Anoplodactylus eroticus*. Measurements were taken from SEM images of specimens in the appropriate position to avoid distorted results. Length was measured on the dorsal surface, from the anteromedial border between chelifores to the posteriormost point on the dorsal surface. Width was measured directly posterior to the chelifores, anterior to the second protonymphon appendage (and the corresponding vestigial bud). Chelifore length was measured from the base of chelifore insertion to the distal pincer tip. Dash indicates missing data.

| Stage | Mean body length (µm) | Mean body width (µm) | Mean cheliforal length (µm) | Features |
|---|---|---|---|---|
| I | 42 | 48 | 55 | protonymphon; chelifores, two pairs of post-cheliforal appendages |
| II | - | - | - | reduction of post-cheliforal appendages |
| III | 110 | 140 | 71 | encysting; primordia of walking legs 1 and 2 |
| IV | 223 | 221 | 142 | encysting; primordia of walking legs 1,2,3, and tailbud |
| V | 456 | 248 | 230 | encysting; limb bud of walking legs 1,2,3; primordia of walking leg 4; eye spots |
| VI | 666 | 248 | 267 | emergent; ocular tubercle |
| VII | 686 | 251 | 392 | juvenile; complete walking legs 1,2,3; limb bud of walking leg 4 |
| VIII | 916 | 317 | 439 | sub-adult; complete walking legs 1-4; tailbud directed dorsally |
| IX | 1444 | 333 | 833 | adult; complete ovigers and genital spurs (male); proboscis protuberances (female) |

*General overview of post-embryonic development*

During copulation, eggs are fertilized externally and transferred to the male. Fertilized eggs are approximately 40 μm in diameter. The male carries embryos in bundles consisting of twenty to fifty embryos each, hung like purses over on modified appendages, termed ovigers, located ventrally on the cephalic segment between the palps and the first pair of walking legs[27]. Protonymphae used their large chelifores to break open the chorion. Once the protonymphon hatches, it grasps onto the outer surface of the egg masses or the male cuticle (**Figure 1A**). In the laboratory this appeared to occur for approximately 1-3 days or until a hydroid was introduced to the petri dish.

Stage I protonymphae were harvested from embryos separated from males in the laboratory (**Figure 1B**). Once separated from adults and egg masses, protonymphae used all appendages to move through water. When the hydroid *Pennaria disticha* was added to the dish, protonymphae immediately moved towards the hydroid, clinging and tearing it with their chelifores. At high densities, the protonymphae destroyed hydroid polyps within a few hours. The protonymphae used their chelifores to assist with entry into the hydroid. When protonymphae density was low, a second stage protonymphon was located. A pinkish substance in the hemolymph of Stage II larvae suggests that the larvae may also be feeding directly on the hydroid.

Stage II is distinguished by the loss of the second and third larval appendages, of which only a stub remains. Stage II protonymphae were collected only in the laboratory, after Stage I protonymphae had been left overnight in a petri dish with the hydroid, *P. disticha*. Thus, the transformation between stage I and stage II appears to occur after the host hydroid has been located.

Subsequent encysting stages (III-VI) were collected within the hydranth of *Pennaria disticha* hydroids attached to substrate in the Kewalo Basin fishing docks. Hydranths bearing encysting pycnogonids were swollen (**Figure 2A**), and encysting larvae were removed with forceps and needles (**Figure 2B**). Encysting stages appeared immobile when collected, however when separated from the hydroid and placed in a petri dish, the proximal cheliforal segments were observed to move laterally, while the outward pincer of the chelate distal portion opened and closed. The encysting protonymphae were not within cysts as has been reported for other encysting larvae. The larval cuticle was transparent, and pink colored hemolymph could be seen within the body cavity, extending into the chelifores and into each anlagen of the adult walking legs, but not the remains of the post-cheliforal protonymphon appendages that later transform into the adult ovigers in *A. eroticus*. In the lab, exuvia was observed being extruded from the polyp (**Video "molt"**). The stage of the shed exuvia could not be determined.

Stage VI larvae were collected both within a swollen hydranth and also during emergence from a destroyed hydranth. Stage VI emerges tail bud first, with the chelifores and proboscis remaining burrowed within the hydranth (**Figure 2C**). At the time of emergence, the cuticle remains transparent and incompletely sclerotized. This cuticle is shed as the Stage VI larva emerges from the hydroid, and the juvenile Stage VII cuticle hardens. In the Stage VII juvenile, three pairs of walking legs are completely developed, including the distal claws (propodus) used to grasp hydroids. The Stage VII juveniles were collected near the hydranth of the hydroids. This stage was always observed feeding on hydroids in the lab, undisturbed by the process of collection and lab preparation. Stage VII 'sub-adults' were located on various sections of the hydroid and appeared similar to adults, yet were smaller in size and lacked secondary sexual characteristics that are fully formed in the adult (Stage IX).

***Postembryonic development in* Anoplodactylus eroticus**

***Stage I (protonymphon, hatching stage)***

*External anatomy described from scanning electron micrographs.*

The first stage *A. eroticus* protonymphon is small, smooth, and spherical, measuring approximately 45 μm in length, width, and height (**Figure 3**). No eye structures detectable. The gut is incomplete, no anus is present. The cuticle is constricted along three lines in all protonymphae examined, superficially dividing the protonymphon into three regions. Although the depth of these grooves may be emphasized due to slight shrinking during specimen preparation, the grooves are positionally consistent in over thirty specimens examined by SEM. In the 'free-swimming' protonymphon, *P. litorale*, Vilpoux and Waloszek (2003)[30] described a single shallow groove on the dorsal surface, the 'abaxial groove'. Here the 'abaxial groove' is renamed the dorsal groove (**Figure 3A**) in order to distinguish it from two additional grooves on the *A. eroticus* protonymphon. The anterior groove extends from the base of each chelifore on the dorsal side, anteriorly, meeting medially at a point between the chelifores, above the insertion of the protonymphon proboscis (**Figure 3B**). The ventral groove divides the ventral surface into two equal sections, the anterior of which includes the proboscis (**Figure 3C**). Over the insertion of the protonymphon proboscis, just posterior to the anterior groove, there is a prominent fold in the cuticle (**Figure 3D**). The protonymphon proboscis is a smooth swelling between and just ventral to the chelifores and differs in shape from the elongate and conical adult proboscis. The mouth, or stomodeum, is positioned medially at the terminal end of the protonymphon proboscis. The stomodeum is surrounded by two flaps which connect on the ventral end and are unconnected and pointed dorso-laterally on the dorsal side (**Figure 3E**). From an anterior vantage point, the stomodeum appears to form the characteristic Y-shape as found in adult pycnogonids, however, a detailed view with SEM reveals this could be an artifact of the shape of a cuticular folds or 'lips' around the stomodeum (**Figure 3E**).

*External anatomy of protonymphon I appendages*

The protonymphon bears three pairs of appendages. The anteriormost first appendage, the chelifore, is by far the largest appendage, approximately equal in length to the larval body (55 μm on average). The chelifore is composed of three segments. The proximal cheliforal segment is inserted anterolaterally, just below the anterior groove, and extends anteriorly (**Figure 3A**). The distal cheliforal segment is directed anteroventrally and terminates in a claw, or chela, composed of an outward movable pincer that curves inwards to meet a fixed pincer. The proximal cheliforal segment moves horizontally and vertically, the distal chelate segments are able to rotate posteriorly. The second and third protonymphon appendages extend laterally. Both second and third appendage pairs are composed of a conical proximal segment (approximately 10 μm) bearing a single setae distally. The medial segment is approximately 25 μm, and distally bears an elongate threadlike portion which divides once, curls, and tapers along its length (**Figure 3A**).

*Internal anatomy based on brightfield Nomarski optics and flourescence microscopy*

Images captured of live specimens using Nomarski optics allowed internal and external structures to be viewed simultaneously. Internally, below the cuticular grooves, the indented epidermis forms points of attachment for striated muscle bundles (**Figure 4 A-C**). Muscles attach at the anterior groove, nearby the region of chelifore insertion (**Figure 4A**). At the anterior groove, two pairs of muscle bundles extend in an posteriorly towards the dorsal groove (**Figure 4 A, B**). Along the dorsal groove, there appears to be four pairs of fibers which radiate from a dorsomedial region laterally and ventrolaterally towards the insertion of the three protonymphon appendages (**Figure 4C**). In this region, the anteriormost fiber divides laterally and one branch connects to the outer base of the chelifores while the other extends to the base of the second appendage. Two fibers run towards the base of the third appendage. The internal anatomy of the protonymphon proboscis was best visualized by using phalloidin as a molecular marker for musculature in conjunction with cLSM (**Figure 5A**). The stomodeum is symmetrically surrounded by six bundles. These fibers fall between (interradial) the tri-radiate fibers (radial) lining the pharynx (**Figure 5A**). Interradial and radial muscles resemble the arrangement of interradial and radial 'lip' fibers observed in the adult pycnogonid proboscis[8]. The intricate structure of the protonymphon proboscis suggests that the larva is able to feed at an early stage.

Cells were identified by the nuclear markers, DAPI and Propidium Iodide. DAPI staining left the specimen in perfect form to be simultaneously observed with brightfield microscopy. Images of specimens stained with Propidium Iodide were captured via cLSM, allowing cellular composition to be assessed in detail by examining a stack of 72 images taken in a dorsal-to-ventral series (at regular intervals < 1 μm depth). Protonymphon neuroanatomy was observed by immunohistochemically labeling nerves with antibodies against acetylated-tubulin and the pan-neural marker, Elav. The expression of Distalless was observed using cross-reactive antibodies generated against Distalless.

At the cellular level, the region between the dorsal and anterior groove is densest in contrast to the posterior region behind the third appendages that contains few cells (**Figure 5B**). The following description of cellular arrangement is organized along the protonymphon dorsal-ventral axis (**Figure 6**; series can be viewed in **Video PI**). Antero-dorsally there are two lobes of cells composed of approximately 15 cells each (**Figure 6A**). On the same plane, positioned approximately where the dorsal suture lies, are about five cells (**Figure 6A**), which correspond to acetylated-tubulin and Elav immunoreactive surface receptors on the protonymphon cuticle (**Figure 6A, 7E**). In the next focal plane is a dense band of cells between and including the cellular aggregations at the base of the chelifores (**Figure 6B**). Moving further ventral though the protonymphon, the anterior dorsal region remains extremely dense consisting of approximately 36 cells in a semi-circular arrangement (**Figure 6C**). Approximately 16 cells are regularly arranged along the outer edge of the semi-circle (**Figure 6D**). In the next ventral focal plane, cells marking the edge of the semi-circle persist, approximately 15 cells comprise the rounded posterior portion and the straight anterior border consists of about 20 cells. The inside of the semi-circle contains no detectable nuclear stain (**Figure 6 E**). This region is strongly immunoreactive to the B-tubulin and Elav antibody (**Figure 6 A, B**), indicating that the anucleate axon tracks of the protocerebral bridge fill the space. In the next more ventral plane, only the anterior border remains between the chelifores, and approximately 16 cells mark the esophagus (**Figure 6 F**). Ventral to the esophagus, two lobes appear consisting of about 15 cells each, posterior of where the pair of lobes appears in the dorsalmost images (**Figure 6 G**). In **Figure 6 H**, there are three anucleate reagions. A medial hollow spot occurs medially in a patch of cells between the second and third appendages and two hollow spots behind the third pair of appendages are surrounded by a ring of cells. Similar to the anucleate anterior region, the gaps correspond to highly immunoreactive areas for axon tract markers (**Figure 7 A, C, F**). Posterior to the insertion of the third appendage pair, cell density is low. From a posterior vantage point, there are a pair of cellular aggregations (**Figure 5 B**). Eyes have not formed but two ocular nerves extend anteriorly from the protocerebrum (**Figure 7 A, B**). Muscles attach at the anterior groove, nearby the region of chelifore insertion (**Figure 4 A**). Distalless immunoreactivity was detected in approximately eight internal cells anterior to the ventral groove and anterolateral to the base of the pharynx, where the commissure of the second and third appendicular ganglia are positioned. Distalless was also strongly detected in cells in the distal cheliforal segment (**Figure 5 C**).

*Internal anatomy of appendages in Protonymphon I*

Within the proximal segment of each chelifore there are two circular, anucleate (**Figure 4 D, 6 F**). One organelle is positioned within the protonymphon body, approximately at the base of the chelifore, and a second pair is positioned directly anterior to the first organelle, within the proximal segment of the chelifore. Meisenheimmer (1902)[47] described the organelles as "glandular cells of the first extremity", and without functional information, there is little that might be added to his initial description. There is no indication that they have an excretion function in the *A. eroticus* protonymphon, as there are no canals connecting the organelles to the ectoderm.

The chelifores are prominently innervated. Each finger in the chela is targeted by a nerve which has divided at its point of origin at the supraoesophageal cheliforal ganglia (**Figure 7 B**). The surface of the chelifores bear sensory receptors (**Figure 7 B**). The proximal and medial segments of each of the two post cheliforal appendages contain regularly distributed cells. The second pair of appendages are innervated by a single nerve leading from the second pair of ganglia heterolateral to the pharynx. The setae located on the proximal segment of the second and third appendages is immunoreactive, and appears to consist of a single nerve cell (**Figure 7 D**). The distal filaments on the second and third appendages are anucleate, and β-tubulin immunoreactivity extends only into the setae between segments

***Stage II: (loss of second and third protonymphon appendages)***

Stage II was attached to neither adults nor hydroids, and was therefore impossible to collect in the field. Only two specimens were identified in the lab and images were captured under the brightfield microscope (**Figure 8**).

The stage II body is distinctly ovoid and filled with a pink colored hemolymph that presumably consists of ingested hydroid tissue. The chelifores remain constantly mobile (**Video 'Stage II'**), yet appear smaller than Stage I, due to disproportionate growth in body size. Post-cheliforal appendages have been reduced to stubs of only the proximal segment, and appear immobile. Eyes are still not externally visible. The cuticle appears to have thickened.

***Stage III (first encysting stage; primordium for first and second pair of adult walking legs)***

The Stage III larva is the earliest stage found within the *P. disticha* hydranth. The third stage remains spherical and has nearly tripled in size; an average specimen measures approximately 140 μm in diameter (**Figure 9A**). The chelifores are directed anteroventrally, however the chela do not curve around the body towards the proboscis as they had in earlier stages (**Figure 9B**). The chela remains articulated at the base of the external pincer as in Stage I and II. The cuticle, blanketing the body, obscures any external view of cuticular grooves including indication of chelifore insertion. The proboscis forms a bulge anteroventrally between the chelifores (**Figure 9B**). The stomodeum is hidden within a slit at the tip of the proboscis (**Figure 9A**). A crease in the cuticle is apparent between the chelifores above the proboscis, reminiscent of the anterior fold between the chelifores in the Stage I protonymphon (**Figure 9C**).

A pair of large heterolateral bulges account for approximately one third of body length (**Figure 9A, D**). This bulge represents the primordia of the first adult walking legs. A crease is present at the base of the bulge, and posterior, a second crease below a slight bulge represents the primordium of the second adult walking leg (**Figure 9C, D**). The cuticle is finely wrinkled in some specimens (**Figure 9A-D**) and smooth in others (**Figure 9E**). Larvae were variously fixed in either formaldehyde or Bouin's solution, and the state of the cuticle might merely reflect differences in fixation.

In several Stage III protonymphae examined, a small pair of bumps are present just posterior to the chelifores (**Figure 9D, F**). Based on position, the bumps appear to be relics of the second or third larval appendages. The bumps were not present in all collected larvae. Under brightfield microscopy, pink hemolymph was observed extending into the chelifores and primordia of the first and second adult walking legs (**Figure 9F**).

### *Stage IV (primordia of first, second, and third pair of walking legs; tailbud)*

*External anatomy of Stage IV*

Stage IV larvae ranged in size from 136 μm to 328 μm, indicating that growth occurs without additional morphological transformation during this stage (**Figure 10A, B, 11A**). Three pairs of distinct, heterolateral bulges represent anlagen of the first, second, and third pair of walking legs. Depending on the constriction of the cuticle these anlagen are more (**Figure 10A**) or less (**Figure 10B, C**) apparent. The proboscis has grown longer and is directed anteroventrally (**Figure 10C**). At the distal tip of the proboscis the cuticle is split revealing a triangular slit, point downwards (**Figure 10D**). The chelifores are similar to Stage III and the distal chelate segment remains mobile. Chelae are directed anterolaterally (**Figure 10B, D**). Posteriorly, a region corresponding to the future tail bud is demarcated laterally by the third appendage anlagen, and ventrally by a groove (**Figure 10E**). The eyes are not yet visible externally.

*Internal anatomy of Stage IV*

Under brightfield microscopy, a pair of small protuberances directly posterior to the chelifores, and presumably equivalent to those observed in stage III specimens, are visible beneath the cuticle (**Figure 11B, 12B**) A pink colored hemolymph fills the body cavity, extending into the chelifores, the three limb buds, as well as into a posterior region that will form the tailbud (**Figure 11C, D**). Cellular arrangement was observed in DAPI stained specimens. A paired chain of ventral ganglia corresponding to the limb buds is apparent in ventral views (**Figure 12A**). The ganglia of the yet unformed fourth adult appendages has begun to form, visible as two small aggregations posterior to the other ganglia (**Figure 12A**). Anterior to the three pairs of ventral ganglia are two additional pairs of DAPI-stained spots at the base of the pharynx (**Figure 12A**). On the dorsal side, between the chelifores, is a semi-circular aggregation of cells, presumably the protocerebrum (**Figure 12B**). It is unclear if some of the cells in this region correspond to the future ocular apparatus.

***Stage V (eye spots; elongation of the first, second, and third pair of walking legs; primordia of fourth pair of walking legs)***

Stage V is approximately equal in width to Stage IV (approximately 223 μm), yet nearly doubles in length (between 430 μm to 550 μm). No body segments are externally visible along the dorsal axis (**Figure 13A**). The three pairs of limb primordia have elongated, extend laterally, and curl back towards the ventral body (**Figure 13A, B**). Both cheliforal segments are longer, and are directed anterolaterally (**Figure 13A, B, C**). The appendicular surface is smoothly annulated and segmentation is unclear (**Figure 15B**). Distally the appendages bear a thorn-shaped tip (**Figure 13C**). The larval proboscis has elongated between the chelifores (**Figure 13C, D**). The Y-shaped stomodeum appears as it will in the adult (**Figure 13C**). In a single specimen, the adult proboscis, terminating in the Y-shaped lips, appears to emerge from the larval stomodeum (**Figure 13E**). A trunk bud forms at the posterior end, flanked by buds of the fourth pair of adult walking legs (**Figure 13F**). The anus has not yet formed. In one specimen, there was a significant tear in the cuticle between the chelifores at the anterior crease observed in specimens at stage I and III that could indicate a break point during ecdysis (**Figure 13G**). Also, this specimen is grasping the walking leg bud with its chelifore, as if had been fixed just prior to molting and is using the chelifore to aid in shedding old cuticle (**Figure 13G**). As in earlier stages, some specimens had a pair of vestigial buds directly posterior to the chelifores (**Figure 13F**).

*Internal anatomy of Stage V*

External and internal anatomy, including neuroanatomy, was simultaneously observable under brightfield microscopy with Nomarski optics. Four ocular spots are apparent anteromedially, dorsal of the chelifores (**Figure 14A**). Also in the head, the protocerebrum is a large mass located above the proboscis and pharynx, just posterior to the four ocular spots. Two nerves emerge anterolaterally from either side of the protocerebrum, targeting each chelifore (**Figure 14A**).

An unexpected observation was the presence of chela on post-cheliforal limb primordia (**Figure 14**). In some specimens, claws appeared to be located on the posteriormost limb (**Figure 14B, C**), while in others, the primordium of the second adult walking leg appeared chelate (**Figure 14C**). Extra clawed appendages were not observed in all specimens, however. Many specimens showed signs of having recently shed the claws (**Figure 14A, E**), suggesting that this is a transient state. In images obtained by SEM, no Stage V specimens bore extra clawed appendages. The transient claws may have been lost during the multiple washes involved in SEM preparation.

***Stage VI (emergent; ocular tubercle)***

Stage VI resembles Stage V, yet increases in length (approximately 666 μm) and has undergone significant morphological change (**Figure 15A**). The larval appendages have extended and the distance between them widened, a small bump representing the ocular tubercle has formed on the dorsal anterior region between the chelifores, and signs of external segmentation are now visible on the dorsal surface. The final, fourth pair of adult walking legs remain as buds curved ventrally, beside the tailbud. In one specimen, the cuticle has ripped along the posterior appendage, revealing an intact, and more articulated appendage below (**Figure 15B**), suggesting that the emergent Stage VI larva in **Figure 15B** is prepared to moult into the Stage VII juvenile.

### *Stage VII (juvenile; limb buds of the fourth pair of walking legs)*

The juvenile Stage VII was always found near the hydranth of *P. disticha* (**Figure 2E**). The body is similar in length and width to Stage VI (approximately 248 μm wide and 686 μm in length, from eye tubercle to tailbud base). The cuticle appears segmented between the first-second and second-third pairs of walking legs (**Figure 16A**). There is no sign of external segmentation between the first pair of walking legs and the cephalon (**Figure 16B**) and the third pair of walking legs and the buds of the fourth pair of walking legs (**Figure 16C**). The proximal cheliforal segment is directed anteriorly, and the distal chelate segment ventrally, with the distal outward finger mobile (**Figure 16B**). The proboscis is directed anteriorly and terminates in a Y-shaped stomodeum (**Figure 16B**). As the in *A. eroticus* adults, palps are absent. However, the cephalic segment is swollen lateral to the proboscis in some specimens (**Figure 16B**). Presumably the swelling represents anlagen of the male ovigers. The tail bud has separated from the posterior limb buds and is directed ventrally (**Figure 16C**). The tail bud terminates with a vertical slit, the proctodeum (**Figure 16C**). The posterior limb buds extend posteriorly and are not externally segmented (**Figure 16C**). The three other pairs of walking legs appear fully formed (approximately 950 μm in length) and are equipped with seven segments, including the distal clawed  tarsi, the propudus (**Figure 16A, D**). As in the adult, slight setae occur on all leg segments (**Figure 16A, D**). The cuticle of Stage VII has thickened, unlike the transluscent, thin cuticle found in encysting stages. Almost immediately particles and epizoans attach to the outer surface of the cuticle (**Figure 16B**).

One specimen appears to have been collected during molting (**Figure 16D**). The cuticle is raised as a sheet over the ventral portion of the body spanning from around the proboscis to the tail bud (**Figure 16D**).

### *Internal anatomy of Stage VII*

Neuronal staining against acetylated tubulin revealed a ladder-like CNS consisting of paired ventral ganglia targeting each pair of walking legs with two prominent nerves (**Figure 16E**). Anterior to the four pairs of ganglia corresponding the four pairs of  walking legs is a pair of ganglia whose target is difficult to distinguish. Presumably they will innervate the ovigers and perhaps palps, if present. The pharynx and tailbud are also highly innervated. The dorsal CNS (brain) is not visible: Thorough staining at this stage is diffiuclt due to variable tissue penetration by the antibody.

### Stage VIII (sub-adult; four pairs of complete walking legs)

Stage VIII sub-adults were collected on different sections of the hydroid, including the base and scape. They were recognized by their small size relative to adults (ranging 650 μm to 1200 μm), but otherwise they appeared similar to the adult pycnogonids bearing four pairs of completely formed walking legs (**Figure 17A, B**). The posterior tail bud is directed dorsally (**Figure 17B**). The ocular tubercle has developed into the characteristic cone-like adult morphology bearing the four eyespots (**Figure 17C**). The proboscis is in its final orientation, directed anteroventrally, inserting ventral to the chelifores on the cephalic segment (**Figure 17B**). The proboscis terminates with a Y-shaped stomodeum (**Figure 17D**).

Adult secondary sexual characteristics were absent or incompletely developed. Chelifore size is similar to the juvenile stage, relative to body length. The legs have elongated relative to body size. The walking legs are clearly articulated (**Figure 17A**) and consist of nine segments: three proximal coxae, a femur, two tibiae, a tarsus, a propudus, and a terminal claw[8] (**Figure 17B**). As in the adults, each thoracic segment bears a prominent lateral process with which the legs articulate[8]. The first pair of walking legs are fused with the cephalon. Similar to the *A. eroticus* adult, short and slight setae are present on the leg segments. Genital spurs and the cement gland tubes characteristic of *A. eroticus* males are absent.

In Stage VIII, separate sexes can be readily distinguished by the presence of ovigerous buds in the males (**Figure 17E**). At this stage the early ovigers consist of a proximal segment, and a smooth elongate portion, slightly annulated, which curves upon itself (**Figure 17E**). The *A. eroticus* female counterpart has no vestigial bumps (vestiges of postchelifloral protonymphon appendages) (**Figure 17A, D**). Also, the sub-female does not bear the characteristic protuberances on the ventral proboscis characteristic of female adult *A.eroticus*[48]. Genital pores were not observed in either sex.

### Stage IX (sexually mature adult)

Adults match the original description of the male *A. eroticus* by Jan H. Stock[49] in 1968. Diagnostic characters are the presence of long genital spurs on the coxae of *all* legs, the presence of short spines on the lateral processes (**Figure 1, 18A**), short but distally erect cement gland tube and the absence of a femoral spur (**Figure 1, 18A**). The female *A. eroticus* was described recently from the Kewalo Basin population[48]. The *A. eroticus* female is characterized by the presence of two pairs of protuberances located ventrally on the proboscis[48] (**Figure 18B**). Palps are absent in both sexes, and ovigers are present only in the male. The legs have genital openings on their second coxae.

**Discussion**

***Life history and ecology of* Anoplodactylus eroticus**

Throughout the year, *Anoplodactylus eroticus* stages were found encysting or living on the hydroid *Pennaria disticha*. Exact time from egg to adult is unknown since encysting larvae could not be cultured long term. The duration of the post-embryonic period may vary based on the time of encystment. For example, Stage II larvae appeared to be induced only after *P. disticha* was introduced to a dish containing Stage I protonymphae. Based on what is known in other encysting species[27,36,50], the post-embryonic period may be is shorter for *A. eroticus* than for free-swimming developers such as *Propallene longiceps*[32] and *Pycngonum litorale*[30,31] which take five months to a year from egg to adult, respectively. Other species of encysting pycnogonids have been reported to have relatively short developmental times, ranging from 21 days[27] to approximately five weeks[36,50]. Lovely[27] attributed briefer post-embryonic development in encysting developers to seasonally ephemeral host hydroids. In the case of *A. eroticus* however, the host hydroid, *P. disticha* does not appear to be seasonally affected.

Still, shorter host life could be a factor. Encysting larvae exert pressure on the host by feeding on host tissue, based on the pinkish substance within larval guts. If encysting pycnogonids are slowly killing the hydranth, they must finish morphogenesis and emerge in a timely manner.

Observations on other encysting species suggest that species-specific associations between pycnogonids and hydroids is species specific[35,51], and therefore information on the relationship between cnidarian hosts and encysting species can prove to be a valuable tool in elucidating details of pycnogonid life history and biogeography. The hydroid *P. disticha* occurs in warm-water seas worldwide. In 1933 *P. disticha* was first reported from Pearl Harbor and Kaneohe Bay, Oahu, found attached to boat bottoms and buoys at old fishing wharves[52]. Prior to recent collections of *A. eroticus*[17,53] only two male specimens of *A. eroticus* had been previously described by Stock (1968)[49]: The holotype was collected in 1951 in the Gulf of Manaar, India, and the paratype was found in 1945 at Honolulu Harbor, Hawaii. Therefore, *A. eroticus* might have either hitchhiked a transoceanic trip to Hawaii on *P. disticha* or otherwise has undergone a host-switch.

*The protonymphon stage of* **Anoplodactylus eroticus** *among the pycnogonids*

As adults, pycnogonids appear to comprise a relatively homogenous taxon. Excepting size and the presence or absence of palps, chelifores, ovigers, or an occasional fifth or sixth pair of walking legs, sea spiders share the same basic features. Likewise, protonymphon (Stage I) morphology is conserved between groups. The thorough description given by Vilpoux and Waloszek[30] of *P. litorale* protonymphon Stage I, juvenile Stage VI, and the adult are quite similar to what we have observed for *Anoplodactylus eroticus*, and we refer readers to their manuscript for additional description.

However, there are certain differences in protonymphae that corresponding to mode of development (e.g. encysting, free-swimming). Like other members of the encysting *Phoxichilidium* and *Anoplodactylus*[27,29,35,36,38], the *A. eroticus* protonymphon is characterized by elongated distal filaments on the second and third larval appendages, lack of spinnerates on the chelifores, and small size (< 0.05 mm). Rather than cling to a substrate, encysting protonymphae are initially suspended in the water column long enough to find a hydroid, lose unnecessary adornments (such as filaments), and encyst.

One difference between the protonymphae described in this investigation and previous reports of protonymphae is the presence of grooves across the body (**Figure 3-3**). Previously the protonymphon has been described a showing no external (or internal) segmentation[30], but this investigation provides evidence that the protonymphon may be clearly segmented. The *A. erototicus* protonymphon cuticle is consistently marked by three grooves. These grooves provide sights for muscle attachment, and between each groove are a pair of ganglia corresponding to a pair of appendages.

*Post embryonic development of* **Anoplodactylus eroticus** *among pycnogonids*

Consistent with prior pycnogonid investigations[30], growth between successive molts is not incremental and a range of sizes was recorded in stages that were otherwise morphologically identical. The thin and flexible encysting cuticle appears to allow for growth without ecdysis. For example, in this report it is possible that Stage V and Stage VI are of the same instar, implying that dramatic growth and morphological change such as the formation of the eye tubercle may have occurred without molting.

Excepting Nakamura (1981)[32], most studies have been unable to directly document the timing of stages by observing ecdysis, and therefore, stage comparisons between pycnogonid species should be made with caution. For example, Stage III in encysting *P. tubulariae*[27] appears equivalent to Stage IV in *A. eroticus,* and likewise Stage IV in *P. tubularaiae* matches *A. eroticus* Stage V. Furthermore, molting does not necessarily reflect morphogenetic processes[54], so stages designated by morphological change should be regarded as distinct from those designated by cuticular molting.

According to Bain's (2003)[5] review of pycnogonid development, encysted larvae pass through a reduced number of stages as compared to species with other modes of development. This investigation finds no support for the idea that development is abbreviated in encysting pycnogonids. Free-swimming developers have been reported to pass through nine (represented in *Achelia alaskensis*[28] and *Pycnogonum littorale*[29]) to fourteen stages (represented in *Pycnogonid littorale*[30,31]) from hatching protonymphon to adult. Like Dogiel's (1913)[29] report of a related encysting species, *Anoplodactylus petiolatus* and *Anoplodactylus pygmaeus* in 1913, this investigation reports 8 stages in *A. eroticus,* not including the mature adult. Six stages until the adult are recognized in the encysting *P. tubularaiae*[27]. Seven stages are recognized in the encysting *Ammothea alaskensis*[28]. Collecting encysting pycnogonids is nontrivial and many investigators might have simply failed to find intermediate stages, leading to an underestimate of transitional stages.

As with the protonymphon stage, certain differences between larval stages in encysting versus non-encysting pycnogonids can be attributed to adaptations benefiting particular life histories, such as cuticular sclerotization, loss of larval appendages, and dynamics of limb growth. In the following discussion, *A. eroticus* (Phoxichiliidae) is compared to the free-swimming *P. littorale* (Pycnogonidae). Phoxichilidiidae and Pycnogonidae are putative sister taxa[53], and importantly, their postembryonic development has been documented in a comparable manner[30].

Owing to protection provided by the host hydroid, the cuticle of encysting *A. eroticus* is transparent, soft, and bears no indication of special structures that might be beneficial for protection or sensory perception. In comparison, the cuticle of free-swimming *P. littorale* is completely sclerotized and bears both setae and smooth patches ("fields") that are putatively used as sensory structures[30]. The Stage II *A. eroticus* loses post-cheliforal appendages, which remain only as small vestigial buds, prior to burrowing within the hydroid, whereas free-swimming Stage II *P. littorale* retains both pairs of post-cheliforal appendages. These appendages likely aid in suspension of *P. littorale* in the water column, whereas in *A. eroticus,* extraneous appendages would only hinder burrowing attempts. The loss of larval appendages prior to encysting is consistent with reports from other encysting species[27,55]. Finally, the shape of the larval stomodeum varies between *A. eroticus* (similar to the adult, a triangular slit surrounded by three lips in the form of a Y) and *P. litorale* (trumpet-like shape, unlike the adult Y-shaped stomodeum). The structural differences in the stomodeum between stages of *P. litorale* as opposed to the uniform structure in *A. eroticus* might reflect the change of host prey between larval and adult stages during the *P. litorale* life cycle.

Other differences between encysting and free-swimming postembryonic development involve the dynamics of limb growth. As a consequence of having to survive in the turbulent marine environment, it is advantageous for certain features of free-swimmers to become functional sooner than in protected encysting developers. In free-swimming larvae, at each molt, one leg pair extends into a long appendage. Presumably as each appendage appears it can function to grasp solid objects, such as hydroid stalks. In encysting larvae, the A-P body axis is patterned prior to elongation, primordia for the three pairs of walking legs and tail bud form prior to extension of the structures. Distal extension occurs simultaneously at a later stage. Once the precursors to the walking legs extend from limb primordia, they are similar in appearance between *P. litorale* and *A. eroticus*. The extended larval appendages are circular in diameter and tubular, they taper distally into ventrally curved and pointed tips, and they are smooth and without segment boundaries. The difference is that in *A. eroticus* Stage IV, for example, all three pairs of walking limb buds appear, while in *P. litorale* the first pair of walking legs are extended and articulated while only a bud of the second pair have formed. The proctodeum appears earlier in *P. littorale* (Stage V) than in *A. eroticus* (Stage VI). This may be a mechanism to prevent the encysting larva from polluting its immediate encapsulation with waste.

Initially it appeared that eye formation was an exception to this trend. The eyes of *A. eroticus* form in Stage V, while the eyes of *P. littorale* have been reported to form at Stage VI[30]. However, Vilpoux and Waloszek based observations on SEM alone. In our SEM of *A. eroticus*, the eyes in Stage V are not apparent as a raised surface. Instead, they are only visible in images captured with brightfield microscopy. Thus, this detail might have been an error based on using only SEM for observations.

***Ancestral mode of post-embryonic development in Pycnogonida***

Based on the presence of a protonymphon in most lineages, the protonymphon can be assumed to have been present in the pycnogonid ground plan. As for post-embryonic development, the ancestral mode is not so clear. According to recent morphological[53] and molecular[56] phylogenetic analyses of pycnogonids, encysting post-embryonic development occurs, at minimum, in three independent clades (Phoxichilidiidae, Tanystylidae, Ammotheidae). Based on variation between clades of encysting developers, this mode of development has evolved more than once. The post-embryonic development of *Ammothea alaskensis*[37] (Ammotheidae) that encysts within the hydromedusae of *Polyochris carafutoensis* more closely resembles that of a free-swimming developer such as *Pycnogonum littorale* than it does the encysting development of *A. eroticus*. Determining the most parsimonious solution based on phylogeny is complicated by the wide distribution of lineages containing encysting species as well as a lack of information on pycnogonid development.

Fossil data supports an early appearance of the free-swimming mode of development with well preserved larvae described from the Upper Cambrian (about 490 Myr ago)[26]. Yet, fossil remains of encysting larvae may be near to impossible to encounter. Vilpoux and Waloszek[30] described development in *P. litorale* as anamorphic to mean that at each larval stage a single additional segment is added. A number of studies have assumed this to be the case for pycnogonids in general, that posterior segments are added sequentially in a series of molts[5,30,57]. In this investigation, it appears that the segmentation process was not definitively anamorphic. Vilpoux and Waloszek[30] predicted that the free-swimming mode of anamorphic development was ancestral based on hypothesized anamorphic development in the arthropod ground pattern (present in the arthropod stem species).

However, arthropod ground patterns have been profoundly influenced by the Articulata concept in which annelids and arthropods are hypothesized to share a recent common ancestor. In the hypothetical arthropod-annelid ancestor, segments originate anamorphically through the proliferative activity of a posterior budding zone as in annelids and certain arthropods[58]. The Articulata concept has fallen out of favor in exchange for the Ecdysozoa concept that separates the annelids from arthropods. In turn, predictions based on the hypothetical arthropod-annelid ancestor must be reassessed. At the present time, the question of the ancestral mode of development in both pycnogonids and arthropods must remain open.

### Comments on a pycnogonid germ band stage

Growth of the encysting larvae is not strictly anamorphic, yet is excluded from a strict definition of epimorphic development as well. For example, the Stage IV larva bears three pairs of limb primordia, each at an equivalent stage of development along the A-P axis (**Figure 3-12**). With the initially differentiated limb primordia and corresponding chain of paired ventral ganglia, this stage is reminiscent of a typical arthropod germband stage embryo. The definition of germband and epimorphic development specifies that segments are formed during embryogenesis, prior to hatching.

Thus, the terms germband and epimorphic development cannot be applied to encysting larval stages – unless the definition were expanded to include post-embryonic development. The significance of hatching as a reference point in ontogenetic studies has been questioned[54], particularly with regard to aquatic larvae which often hatch as three or four segmented head-like larvae. In crustaceans, it has long been known that the naupliar and post-naupliar segments are patterned at different times during development. On a molecular level, the process of segmentation in the head is different from that in the trunk, even in arthropods which pattern head and trunk regions synchronously[59]. Thus, although segments are added after hatching, body patterning in encysting pycnogonids is closer to the epimorphic, than the anamorphic side of the gradient.

Understanding the dynamics of body patterning is important in light of recent studies in arthropod evolution and development which use gene expression as molecular markers for segmental homologues. Gene expression has been shown to be spatially dynamic over time. In order to interpret gene expression patterns it is important that the taxa being compared are at corresponding stages. Generally, the point of comparison during arthropod development is the germ band stage in which the basic organization of the of the body is initially differentiated. Like the aquatic naupliar larvae of crustaceans, pycnogonids do not display an embryonic germ band stage. The anterior "head" segments have already differentiated completely by the time the protonymphon hatches, in contrast to the more posterior segments that form in the encysting stages. Although in *A. eroticus* the initial anterior segmentation phase has been decoupled from the phase of body patterning at Stage IV, initial differentiation along the A-P axis is clearly discernable at Stage IV and resembles the typical arthropod germband. Likewise, the development of the "post-naupliar germ band" has been described in various malacostracan crustaceans[41,60,61].

Following this precedent, Stage III to IV might be termed the post-protonymphon germ band. The question is; how different are these stages from embryonic germ band, and can embryonic and post-embryonic germ band be used as equivalent stages in comparative studies?

### *The pycnogonid head*

A central problem with the pycnogonid head is that it has yet to be consistently defined. Some maintain that the pycnogonid head includes the first pair of walking legs due to an anteriorized fusion of this segment in the adults of most pycnogonids[30]. I disagree based on observations during development and instead favor a traditional interpretation[12] that the pycnogonid head is the anterior region including the proboscis and the first three segments bearing chelifores, palps, and ovigers (when present), and *not* the first pair of walking legs. The walking legs are distinguished from head appendages for the following reasons. In *A. eroticus*, the formation of the walking legs is distinct from that of anteriormost three appendages, that have differentiated prior to hatching. As an adult, walking legs are nearly identical to one another along the length of the body and distinct from the three anteriormost appendages.

At Stage V (**Figure 3-15A**) a furrow appears in the cuticle distinguishing the first pair of walking legs from the cephalon. Fusion of the walking legs with the cephalon occurs at some point after the juvenile VII stage, during which the ganglia targeting the first pair of walking legs are distinct from the suboesophageal ganglia targeting ovigers (**Figure 3-16E**). Furthermore, in certain pycnogonid species, the adult cuticle is externally segmented between the ovigers and the first pair of walking legs (e.g. *Nymphon hispidum* and *Nymphon perlucidum*)[12].

Accepting that the pycnogonid head consists of three appendage-bearing segments, and does not include the first pair of walking legs, lends support to the idea that the protonymphon represents the early adult head[8]. An additional reason for including the first pair of walking legs in the pycnogonid head was theory driven: The ancestral arthropod head hypothetically consists of four appendage-bearing segments[58,62,63]. A protonymphon and pycnogonid head bearing three appendage-bearing segments is unique among arthropods. 'Appendage-bearing' segments is a critical distinction – previously the protonymphon has been considered to have four segments based on the presence of four pairs of ganglia, and only three pairs of appendages[17]. Whether a 3 or 4-segmented head represents a primitive state for arthropods in general is left to future phylogenetic analyses.

### *A pycnogonid labrum*

Pycnogonids are the only extant arthropod taxon missing the labrum, a "lip-like" structure above the stomodeum[2,17,64]. This structure unites all other arthropods to the exclusion of pycnogonids. The labrum either; (a) represents a novel structure appearing after the emergence of the pycnogonid lineage, or (b) has been independently lost or unrecognizably transformed in the pycnogonid lineage. In studies of arthropod head evolution, the origin of the arthropod labrum is of particular interest[65-67]. Two current competing hypotheses predict that the labrum either originated as a simple outgrowth over the stomodeum or originated on an anteriormost segment from a pair of fused appendages[40]. In pursuit of an explanation for the pycnogonid equivalent, the dorsal pycnogonid proboscis[16,68] or the chelifores have been implicated[17]. In preliminary analyses of embryonic development (**Appendix 3.1**) and a review of the historical literature, there is no clear evidence linking proboscis development or post-oral protuberances to labrum development. In the proboscis=labrum hypothesis, the dorsal proboscis (upper third portion of the "Y" shape) is considered to be equivalent to the labrum based on the observation of bi-lobate structures in the 'proboscis anlagen', resembling labrum anlagen[15]. In this investigation, bi-lobate cellular aggregations were observed at the base of the protonymphon and larval proboscis (**Figure 3-6A; 3-12B**).

The three sections comprising the proboscis are fused into a Y-shape not only in the protonymphon, but earlier in the embryo prior to hatching. In the protonymphon the pharynx is innervated and muscularized, the stomodeum is open, and the protonymphae appear to feed. There seems no reason to assume the bi-lobate aggregations at the base of a distinct larval proboscis are the adult proboscis anlagen.

The association of the cheliforal ganglia with the protocerebrum led to a hypothesis that chelifores represent appendages associated with the anteriormost segment which have since been lost or transformed into the labrum of other extant arthropods[17]. Immunoreactivity against acetylated tubulin in the protonymphon did corroborate the earlier report[17] showing a pre-oral commissure corresponding to the protocerebrum, connecting cheliforal ganglia (**Figure 3-7A-E**). The association of the cheliforal ganglia with the protocerebrum led to a hypothesis that chelifores represent appendages associated with the anteriormost segment that have since been lost or transformed into the labrum of other extant arthropods[17]. Furthermore, Dll expression can be interpreted as a factor in support of the homology between chelifores and the labrum. Dll expression has been reported in the labrum of a chelicerate, diplopod, insects, and in both directly and indirectly developing crustaceans[69]. This expression pattern has been used to suggest that the labrum is of appendicular origin, as Dll is known to be expressed in nearly all arthropod appendages. In the protonymphon, Dll was expressed as expected in the distal segment of the chelifores. This expression pattern is consistent with both the known role of Dll in outgrowth patterning, as well as with theories of a cheliforal origin of the pycnogonid labrum. However, it would be premature to infer homology based on this data alone. Future research on gene expression in the chelifores will help to elucidate this exciting problem in arthropod evolution.

### *Transient larval appendages*

Lack of Distalless expression in post-cheliforal protonymphon appendages was unexpected. The second and third larval appendages consist of two segments and a distal filament in the protonymphon stage, and only short proximal buds by Stage II. In subsequent larval stages, only a single pair of buds remain. The absence of Dll expression in these appendages is reminiscent of the dynamics of Dll expression in crustacean mandibular appendages. The mandible of certain basal malacostracan crustaceans is composed of two segments, the proximal coxopodite and distal telopodite[42]. In these mandibles, Dll is continuously expressed during development. In other malacostracans, the mandible is missing the telopodite, and Dll is only transiently expressed and subsequently eliminated[42]. Likewise, the lack of Dll expression in the post-cheliforal protonymphon appendages may foreshadow the reduction of this appendage during development.

The transient claws on the limb primordia of Stage V larvae (**Figure 3-14**) provide a reminder that appendicular form may be labile. Prior to the last few decades of research in evolution of development, structures, such as appendages, were considered to be relatively conserved over evolutionary time. In the arthropods this has led to over a century of complicated scenarios aligning appendages along the body axis across taxa based on structural and positional similarity[70: p 200]. Following this precedent, the post-cheliforal claws along the Stage V body axis might be homologized with the uniramous chilaria along the opisthosoma of Xiphosurans (includes the horseshoe crab, *Limulus*). However, the larval claws are lost within the same stage in which they are first observed. Although the underlying molecular basis of claw development has yet to be elucidated, it has been demonstrated that alteration in the expression of a single gene can transform one appendage into another[71,72]. Assigning homology has become all the more complicated.

### *Conclusion*

Embryonic and post-embryonic descriptions of pycnogonid development generated with modern techniques are overdue. Already, the qualitative observations of a three segmented head and simultaneous patterning of body segments in *A. eroticus* development prompt reassessment of popular predictions on the pycnogonid ground pattern such as a four segmented head and anamorphic growth from a posterior budding zone[63]. Once enough primary data is generated, Pycnogonida will prove an important taxon for 'evo-devo' investigations of arthropod evolution. Due to their unique morphology and putative basal phylogenetic status, pycnogonids provide an ideal group for exploring variation within Arthropoda, and could possibly reveal information on the origins of crown group arthropods.

**Acknowledgments**

Many thanks are due to Mark Martindale, Elaine Seaver, William Browne, David Matus, Craig Magie, Heather Marlow, Kevin Pang, Andreas Hejnol, Richard Schalek, and Cassandra Extavour for technical assistance. I thank the staff of Kewalo Marina, Ana Poe, and Neils Hobbs for collection assistance. Constructive comments on the manuscript were kindly provided by James Hanken, William Browne, and Andreas Hejnol.

**Figure 1**

Protonymphon larvae of *Anoplodactylus eroticus* hatch from embryos carried on the ovigers (O) of adult males. **A**. Ventral view of male bearing embryos, chelifores (Ch) directed down. First (left-hand) walking leg was damaged during SEM preparation. **B**. Embryos and a freshly hatched protonymphon, with anterior chelifores (Ch) directed towards the reader.

**Figure 2.**

Images of live *A. eroticus* specimens under dissecting microscope**. A**. Swollen hydranths of the hydroid *Pennaria disticha* indicated *A. eroticus* larvae located within. **B**. Larvae dissected from hydroids with pins. **C**. Stage VI larvae emerge from the hydroid, tailbud first, with the chelifores and proboscis remaining burrowed within the hydranth.

**Figure 3**.

Stage I, protonymphon larva of *Anoplodactylus eroticus*. **A-C**. Chelifores directed anteriorly, and arrows indicate cuticular grooves. **A**. Oblique dorso-lateral view showing dorsal groove. **B**. Dorsal view showing anterior groove. **C**. Ventral view showing ventral groove. **D**. Chelifores directed towards bottom right corner. Oblique dorso-anterior view showing the cuticular fold (asterisk above) located between the chelifores, and between the anterior groove and the proboscis. **E.** Chelifores directed towards reader, anterior view of the stomodeum. Abbreviations: S, stomodeum; Ch, chelifore; A2, second larval appendage; A3, third larval appendage. Oc, ocular apparatus; Ch, chelifore; W1, walking leg 1; W2, walking leg 2; W3, walking leg 3; W4, walking leg 4; T, tailbud.

**Figure 4.**

A single *Anoplodactylus eroticus* protonymphon viewed under Nomarski optics. Chelifores (Ch) directed up. Images captured at four focal planes, from dorsal (**A**) to ventral (**D**), showing muscular bundles with attachment points at the anterior groove (AG; **A**) and dorsal groove (DG; **B, C**). Raw images are in the left column. In the right column, muscles are artificially colored red, anucleate glands in the chelifores (G) are colored yellow.

**Figure 5.**

Fluorescence microscopy of *Anoplodactylus eroticus* protonymphae stained with cross-reactive molecular markers. **A**. Dorso-anterior view. Phalloidin staining (green) reveals six muscular bundles (r, radial fibers) radiating outwards from the protonymphon proboscis (P) and tri-radiate muscular bundles (ir, interradial fibers) surrounding the Y-shaped pharynx (Y). **B**. Dorsal view. Propidium Iodide nuclear marker with cells artificially depth-coded, colors range from warm (red) indicating dorsal to cooler (blue) indicating ventral. Cell density is low in the posterior protonymphon, except for a pair of heterolateral cellular aggregations (arrows). **C**. Distalless (green) was expressed in approximately eight cells within the anterior half of the protonymphon body and in the distal segment of the chelifores. No Dll was detected in protonymphon appendages. Abbreviations: Ch, chelifore; A2, second larval appendage; A3, third larval appendage.

**Figure 6.**

Protonymphon of *Anoplodactylus eroticus*. Cells labeled by Propidium Iodide. Panels were selected from a stack of images captured on the cLSM at 72 focal planes ranging from dorsal (**A**) to ventral (**I**). Darkly colored cells represent the focal plane of interest, faint grey image in the background of each panal is the compressed image of all focal planes.

**Figure 7.**

Neuroanatomy of *A.eroticus* protonymphon visualized by immunostaining of

Elav (**A-D**) and acetylated tubulin (**E-G**). In all panels, anterior chelifores are directed up. **A-D**. In the upper row, ocular nerves exiting the dorsal surface of the protocerebrum are artificially converted from red to yellow, the bottom row is raw data. **A,C.** Dorsal view. **B, D.** Oblique dorso-lateral view. **E**. Dorsal view. **F.** Ventral view showing that A2G and A3G are connected by separate commissures. **G.** Enlargment of larval appendages showing that distal filaments and setae of second and third larval appendages are innervated by a single cell located in the corresponding, proximal segment of insertion (arrows). Abbeviations: CG, cheliforal ganglia; SR, sensory receptor; A2G, ganglia targeting second larval appendage; A3G, ganglia targeting third larval appendage.

**Figure 8**.

Stage II larva of *Anoplodactylus eroticus*. Chelifores (Ch) remain mobile, while only a proximal bud remains of the second (A2) and third (A3) larval appendages.

**Figure 9**.

Stage III encysting larva of *Anoplodactulus eroticus*, characterized by functional chelifores (Ch), and primordia of first (W1) and second (W2) walking legs. **A**. Ventral view, showing slit-like stomodeum (S). **B**. Lateral view. **C**. Dorso-anterior view, showing a crease (asterisk below) between chelifores. **D**. Dorso-lateral view, showing a vestigial bud of the post-cheliforal protonymphon appendages (arrow). **E**. Dorso-lateral view of a smoother and larger Stage III specimen. **F**. Pink hemolymph within larva reveals the extension of gut diverticula into the proximal chelifores, and primordia of W1 and W2. Arrow indicates vestigial bud.

**Figure 10**.

Stage IV encysting larvae of *Anoplodactylus eroticus*, characterized by functional chelifores (Ch), primordia of first (W1), second (W2), third (W2) walking legs, and the early tailbud (T). Chelifores directed up (**A-C**). **A, B**. Ventral view, showing size and cuticular variation at Stage IV. **C**. Ventro-lateral view. **D**. Anterior view, stomodeum (S) opens into a triangular slit. **E**. Posterior view, showing tailbud anlagen (T).

**Figure 11**.

Brightfield images of live Stage IV *Anoplodactylus eroticus*. Chelifores directed up (**A-D**). Pink colored hemolymph in gut diverticula extends into the chelifores (Ch), limb primordia, and early tailbud. **A**. Three Stage IV larvae demonstrate size variation. **B**. Dorsal view, vestigial bud of the post-cheliforal larval appendage indicated by an arrow. **C**. Dorsal view of a different specimen from B, buds absent. **D**. Ventral view.

**Figure 12.**

A single *Anoplodactylus eroticus*, Stage IV encysting larva viewed under Nomarski optics (left), beside the complementary image of the specimen stained with DAPI nuclear marker (right; blue). **A**. Ventral view. DAPI nuclear marker reveals five pairs of ganglia (G1-G5). The first corresponds to vestigial buds of the post-cheliforal larval appendages, three pairs correspond to primordia of walking legs (W1-W3), and the smaller fifth pair will eventually correspond to the fourth pair of walking appendages. **B**. Dorsal view shows anteromedial paired cellular aggregations between the chelifores (arrowhead). In the brightfield image, arrows indicate the bud of a post-cheliforal larval appendage.

**Figure 13**.

Stage V encysting larvae of *Anoplodactylus eroticus*, characterized by functional chelifores (Ch), extended buds of walking limbs (W1, W2, W3), and primordia of the fourth walking appendage (W4) and tailbud (T). Chelifores directed up (**A-C, E,F**). **A**. Dorsal view, arrows indicate a cuticular horizontal furrow between the cephalon and W1. **B**. Ventral view, arrow indicates the thorn-shaped terminal of limb buds. **C**. Dorso-anterior view, showing the Y-shaped stomodeum (S). **D**. Oblique ventro-lateral view. **E**. Ventral view of the cephalon, showing the Y-shaped pharynx emerging from the larval proboscis. **F**. Dorsal view, arrow indicates post-cheliforal bud. **G**. Anterior view, chelifores directed down. A tear in the cuticle (asterisk above) between the chelifores may indicate a breakpoint in the cuticle during ecdysis. Also the chelifore (arrow) is grasping W1, indicating that the specimen may use chelifores for aid in molting.

**Figure 14.**

Stage V encysting larvae of *Anoplodactylus eroticus* bear transient claws (arrows) on limb buds 1-3. Chelifores directed up in all panels. **A**. Dorso-anterior view, showing internal anatomy of first walking leg

(W1) and head including the ocular apparatus (Oc) and protocerebrum (Pr) with two pairs of heterolateral nerves targeting the chelifores (yellow arrowhead). **B**. Ventro-posterior view. **C**. Ventral view. **D**. Ventral view. **E**. Dorsal view, showing shed chelae, as in **A**.

**Figure 15.**

Stage VI ecysting larvae of *Anoplodactylus eroticus*. **A**. Oblique dorsal view. **B**. Dorso-posterior biew, showing breaks in the cuticle (asterisk above) of W3, with intact appendage below. Abbreviations: Ch, chelifore; Ot, ocular tubercle; W1, W2, W3, W4, first, second, third, fourth walking leg (respectively); T, tailbud.

**Figure 16.**

Juvenile (Stage VII) *Anoplodactylus eroticus*, characterized by three pairs of complete walking legs (W1-W3) including the claw-like propudus (P), buds of the fourth pair of walking legs (W4), and inverted tailbud (T). **A**. Ventral view. **B**. Enlargement of head in (A) showing heterolateral swellings at the base of the proboscis (arrow) and an epizoan living on the juvenile cuticle (arrowhead). **C**. Enlargment of posterior region of (A) showing the anus. **D**. Chelifores directed downwards and towards reader. Ventral view of a juvenile undergoing ecdysis, older cuticle remains on the ventral surface (arrow). **E**. Neural structures labeled with anti-acetylated tubulin and artificially depth-coded, colors range from warm (red) indicating dorsal to cooler (blue) indicating ventral. Ladder-like CNS consists of a pair of ventral ganglia targeting each walking leg with two nerves. Arrow indicates a distinct pair of subpharyngeal ganglia anterior to the first pair of walking legs. Pharynx and tailbud are highly innervated. Dorsal neuroanatomy (including the brain) is less visible.

**Figure 17.**

Sub-adult (Stage VIII) *Anoplodactylus eroticus*, characterized by complete set of articulated appendages, four ocular spots (Oc) on the ocular tubercle (Ot), and a dorsal tailbud (T). **A**. Ventral view. **B**. Lateral view. **C**. Oblique dorsal view of cephalon. **D**. Ventro-anterior view of female showing filamentous "teeth" in the Y-shaped stomodeum (S). **E**. Ventral view of male cephalon showing ovigerous buds (O). Abbreviations: Ch, chelifore; L, lateral process; c1, coxa 1; c2, coxa 2; c3, coxa 3; f, femur; t1, tibia 1; t2, tibia 2; ta, tarsus; p, propudus; cw, Claw.

**Figure 18.**

Adult *Anoplodactylus eroticus*. **A**. Ventral-lateral view of male bearing ovigers (O), and legs with a prominent genital spur (G) and short cement gland (C). **B**. Anterior-lateral view of female cephalon bearing two pairs of ventral protuberances on the proboscis (arrowheads). Stomodeaum (S) is open, forming a triangular shape.

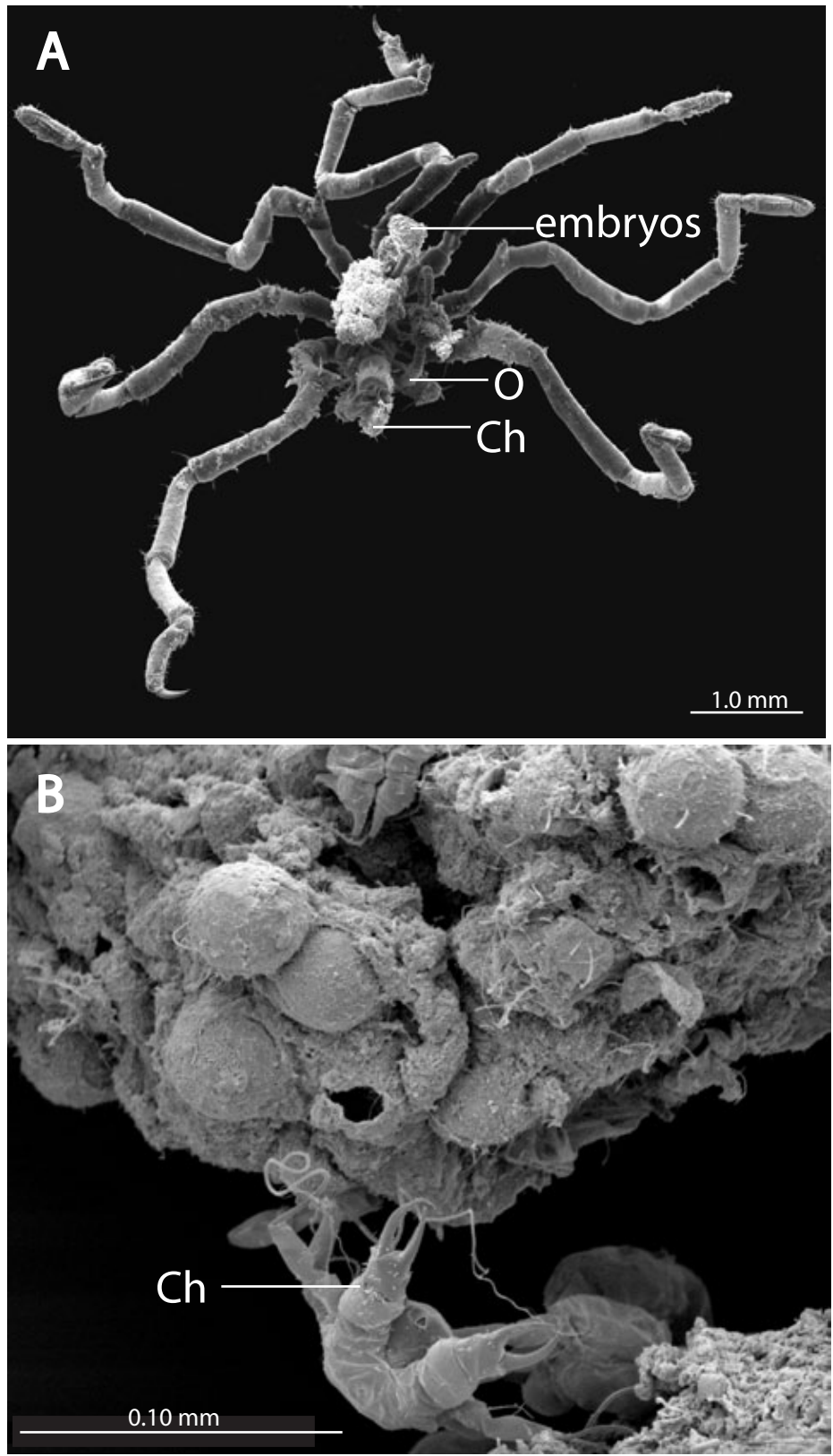

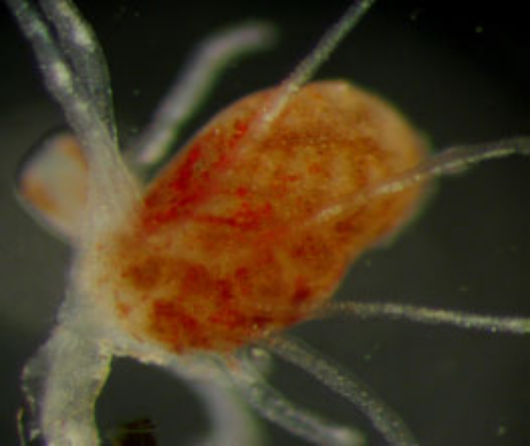
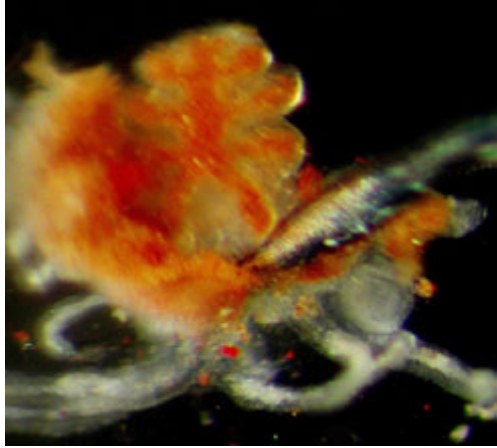
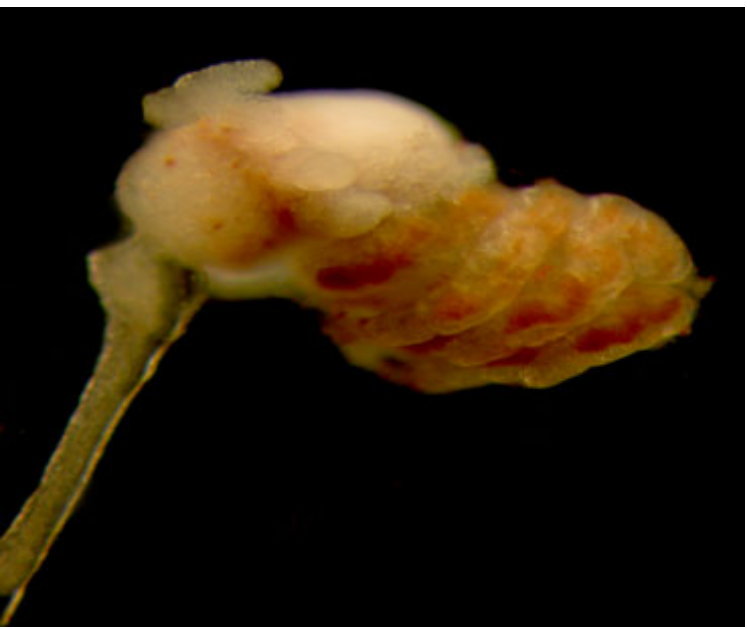

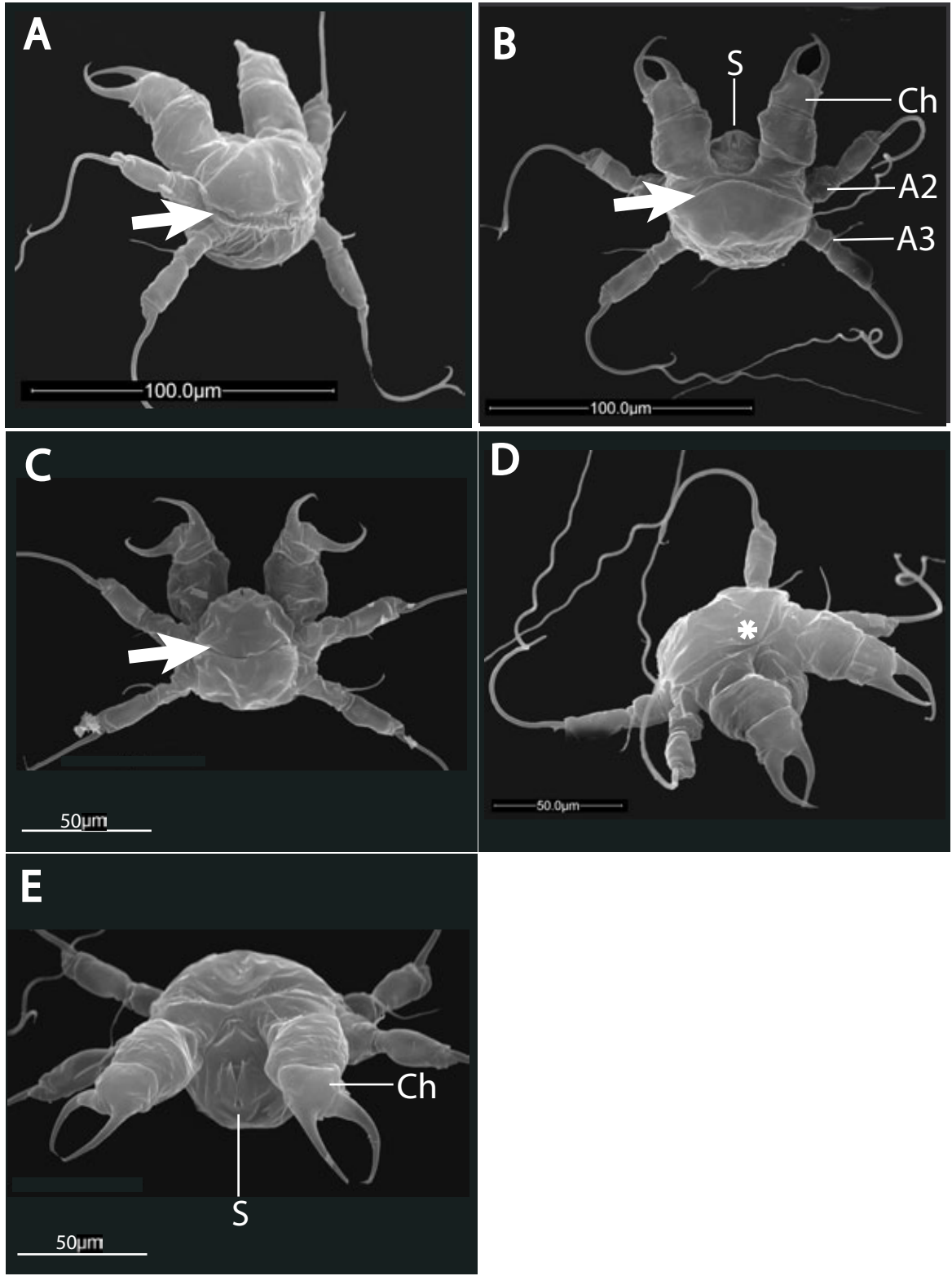

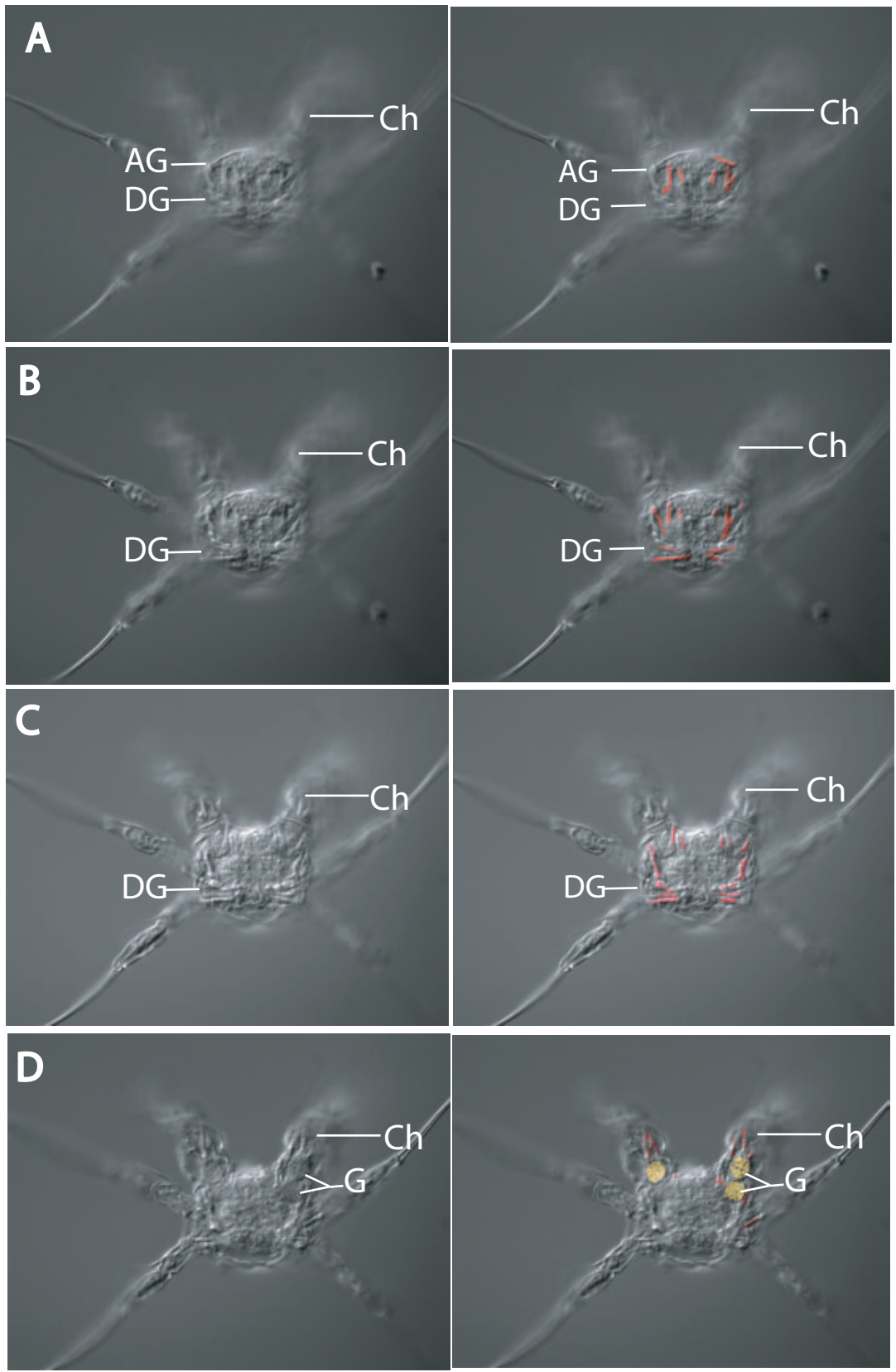

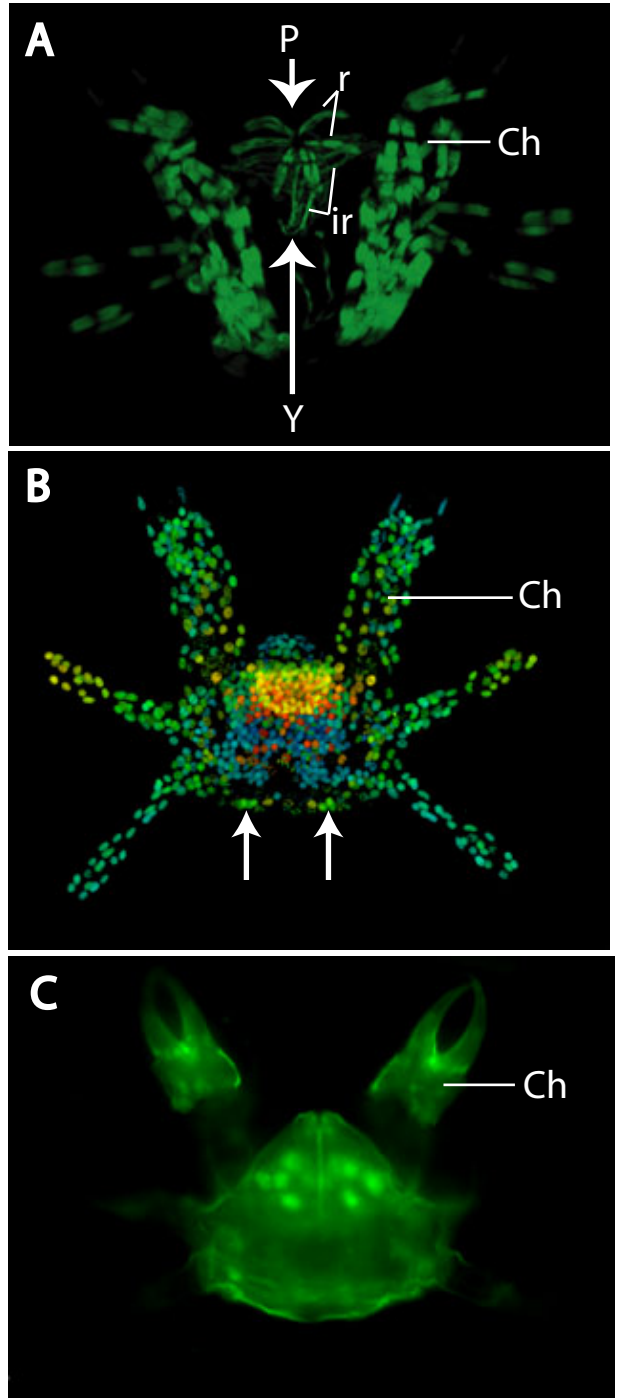

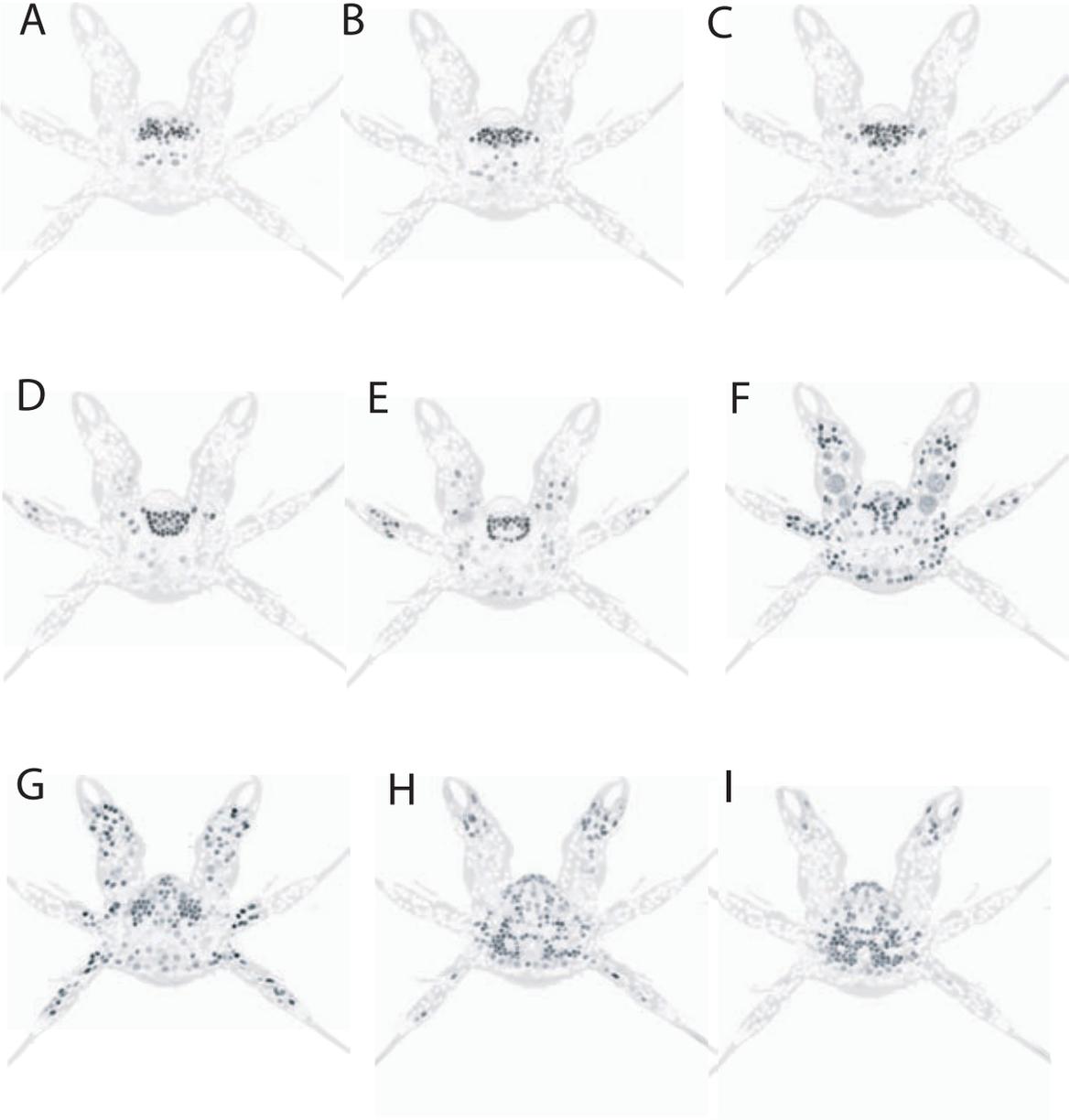

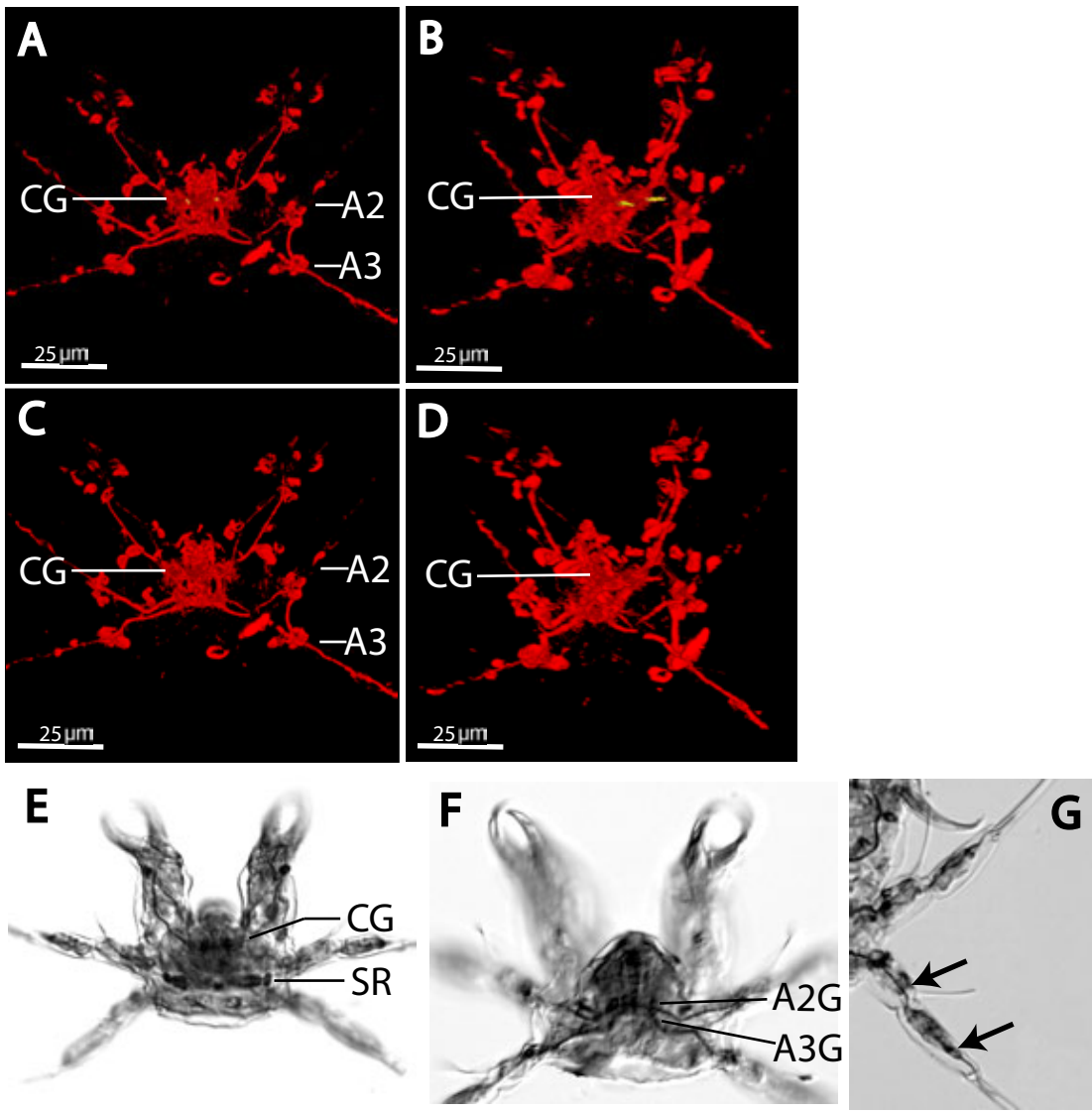

Figure 3-8 (Continued)

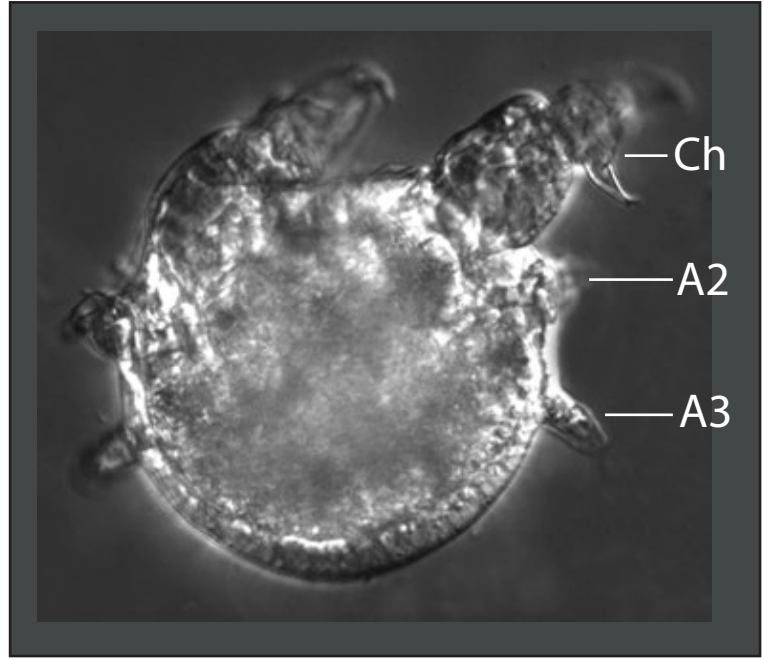

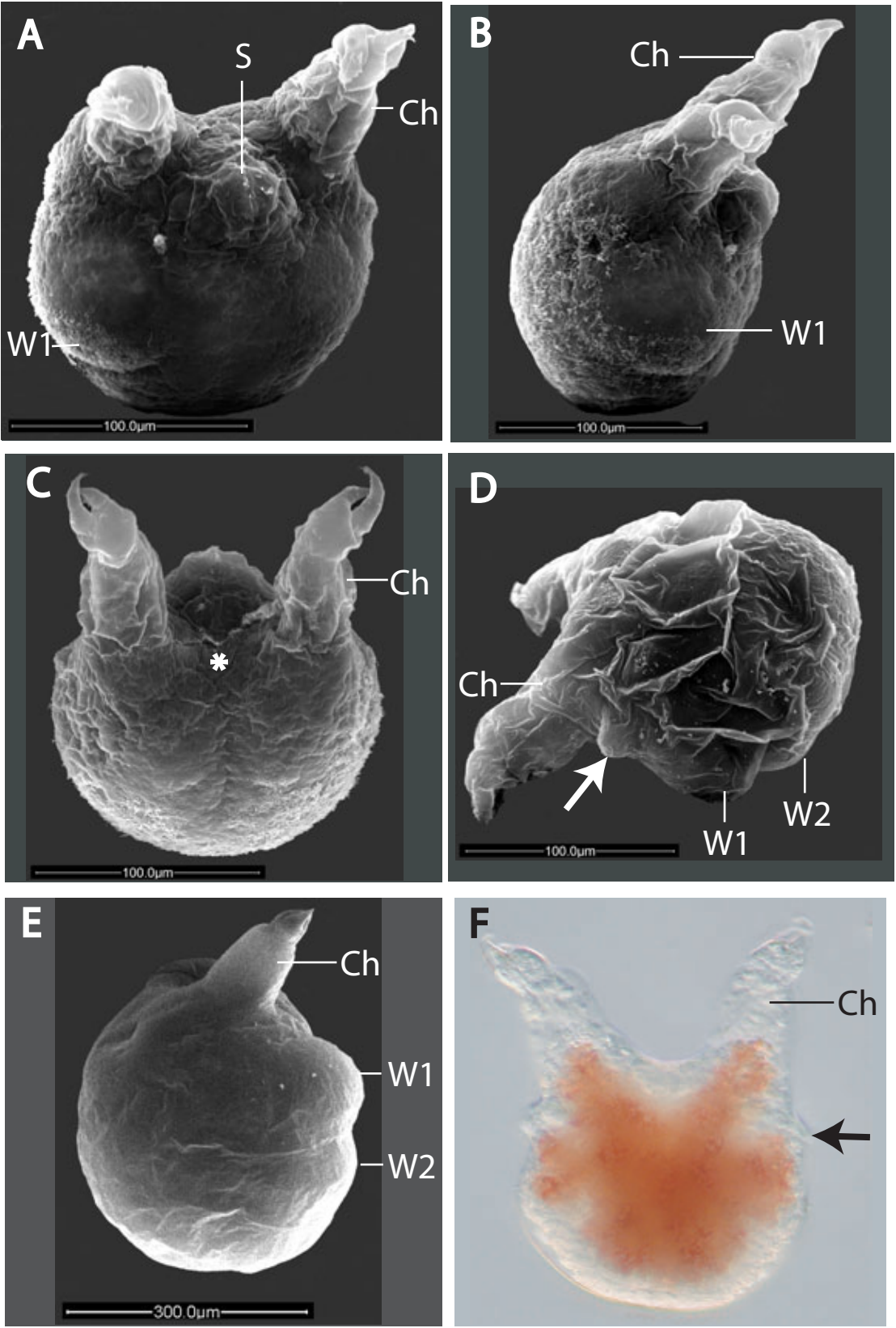

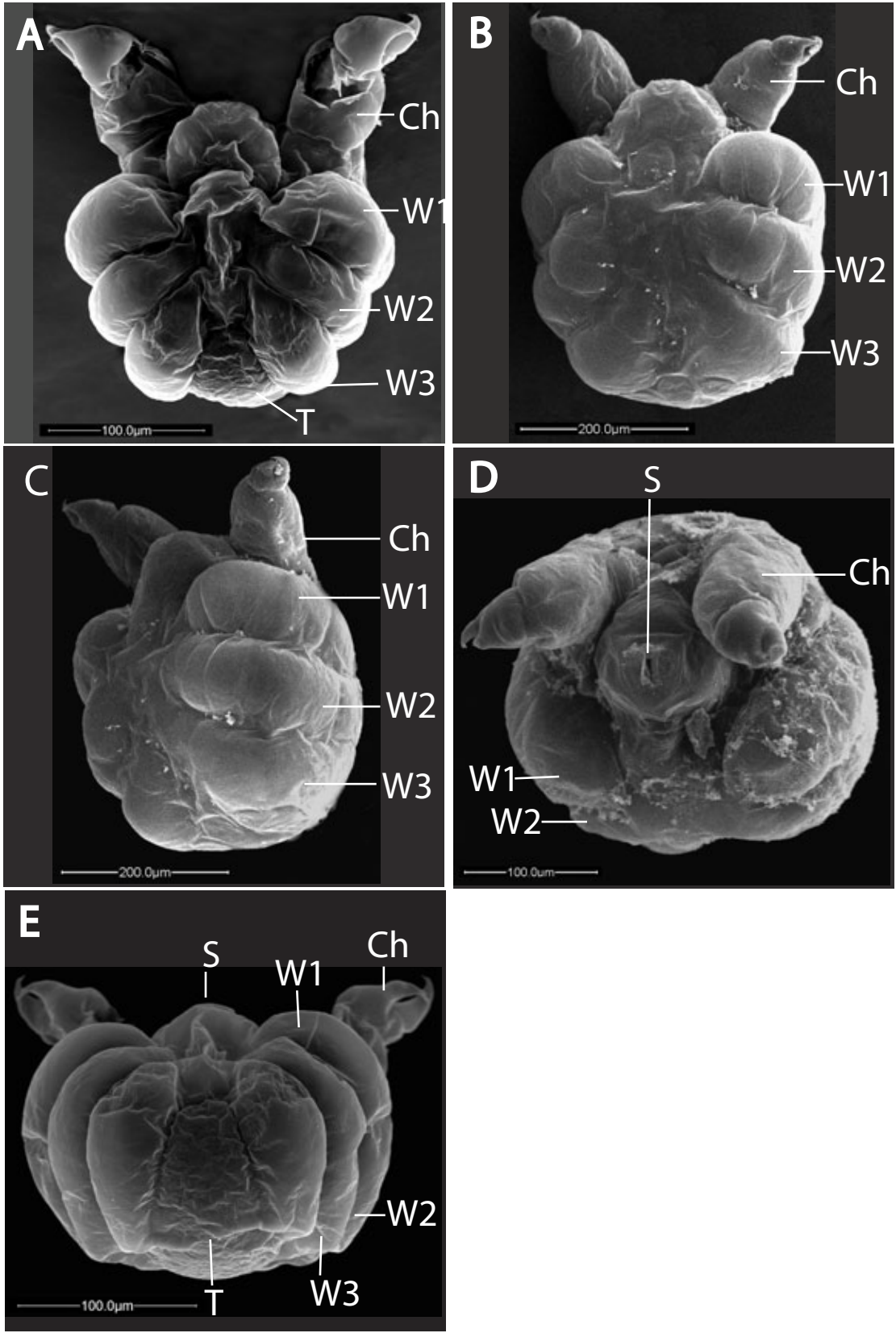

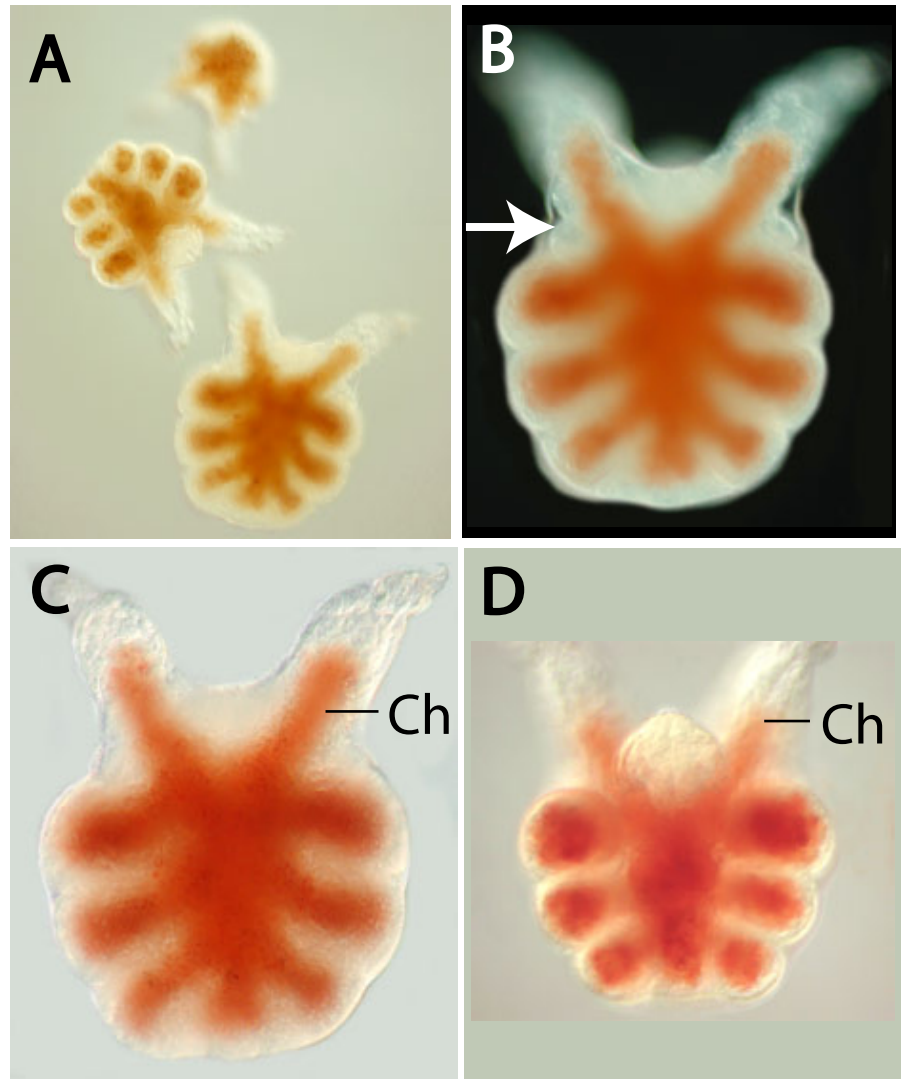

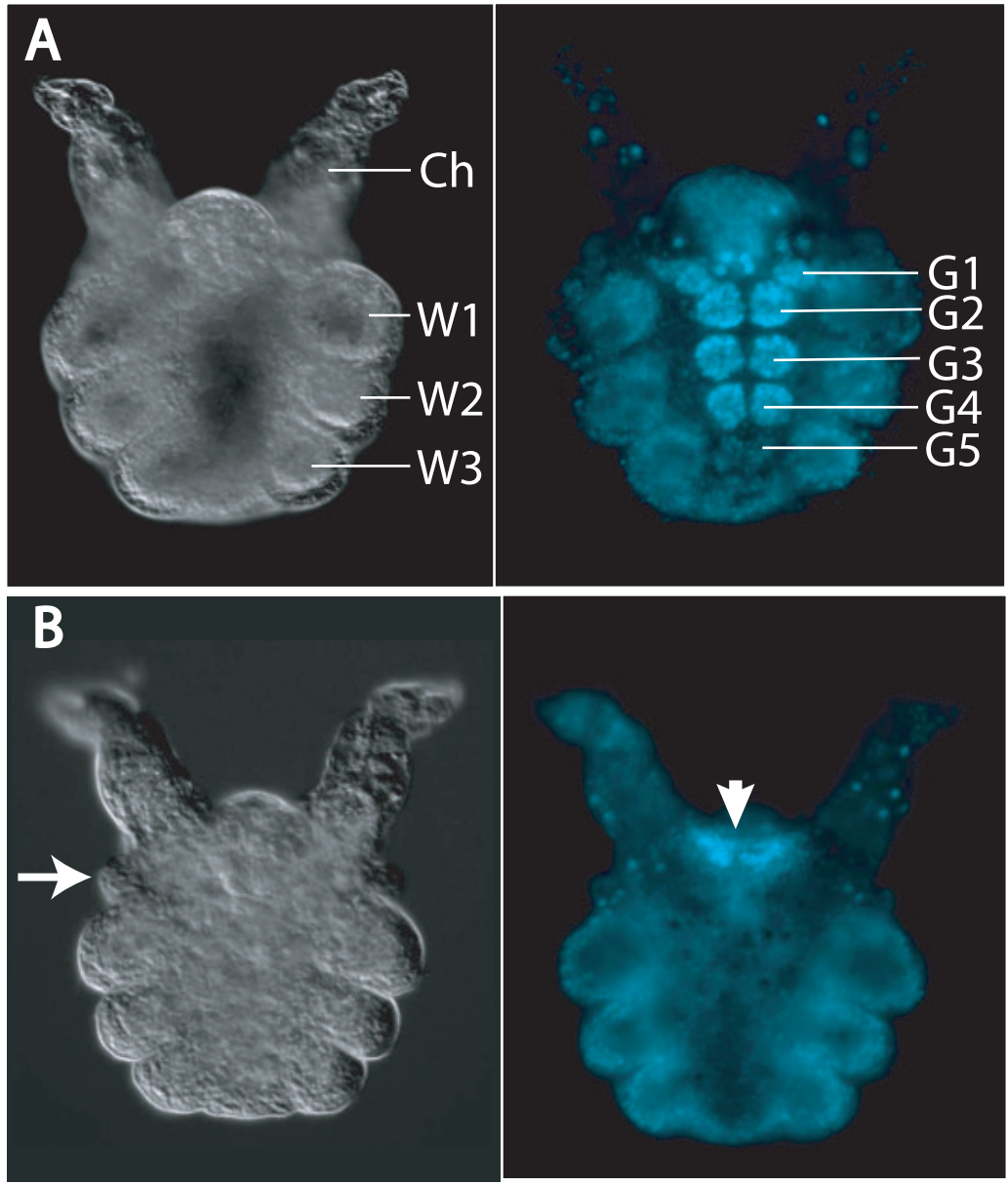

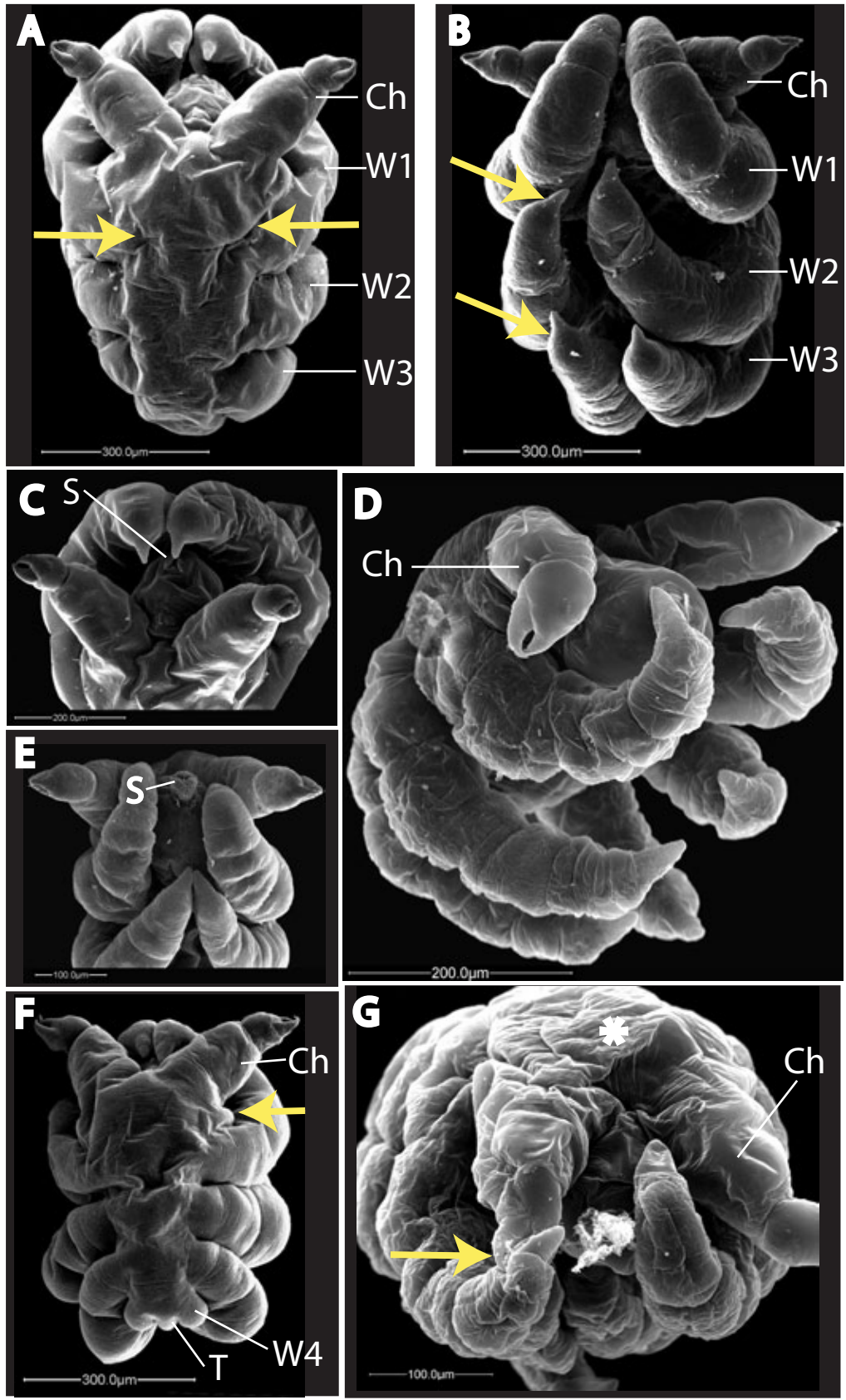

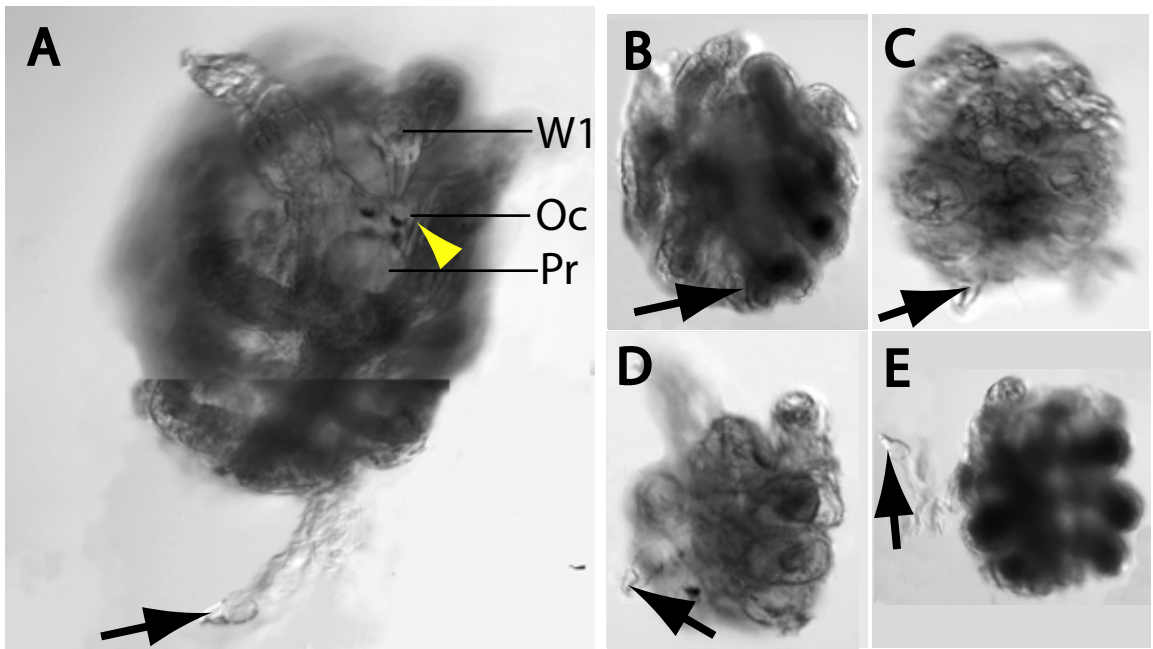

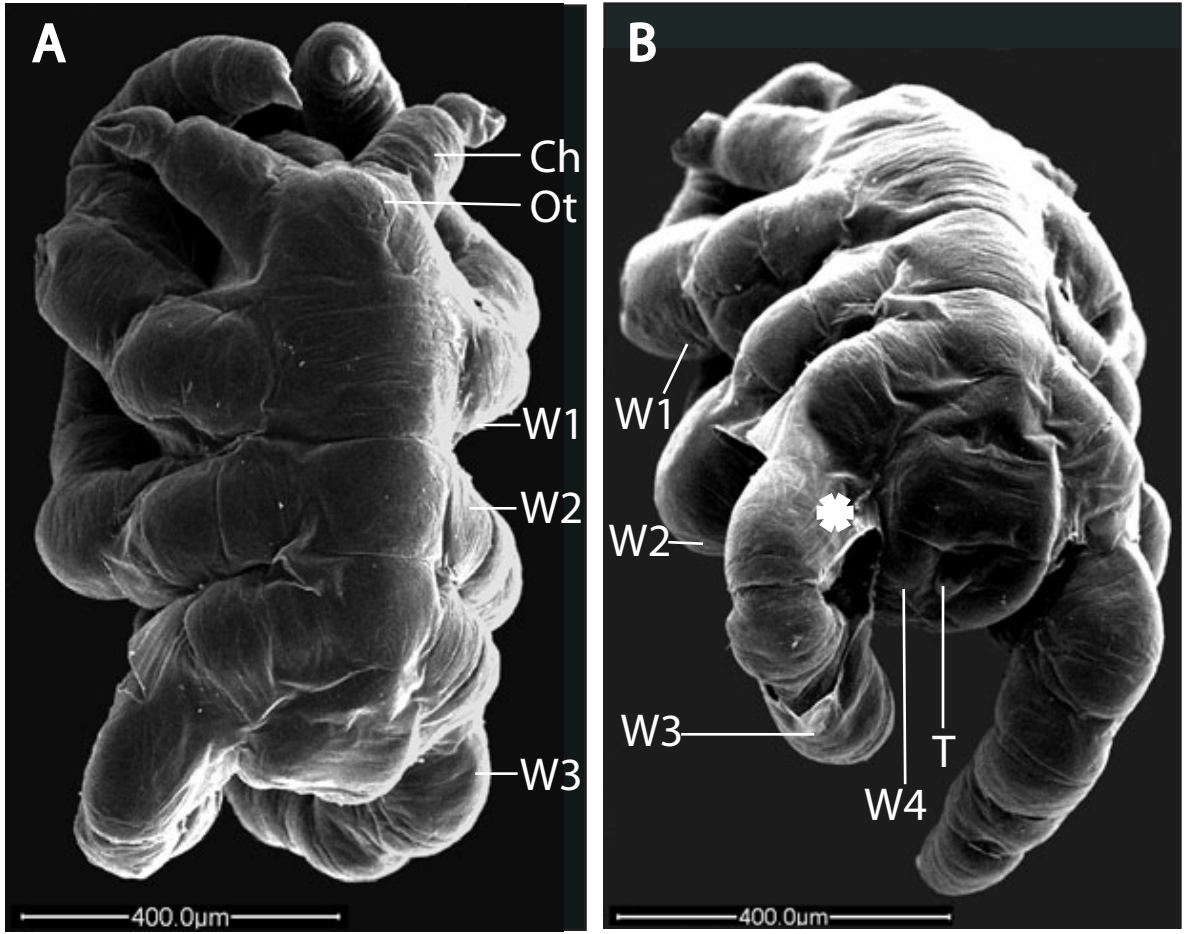

Figure 3-16 (Continued)

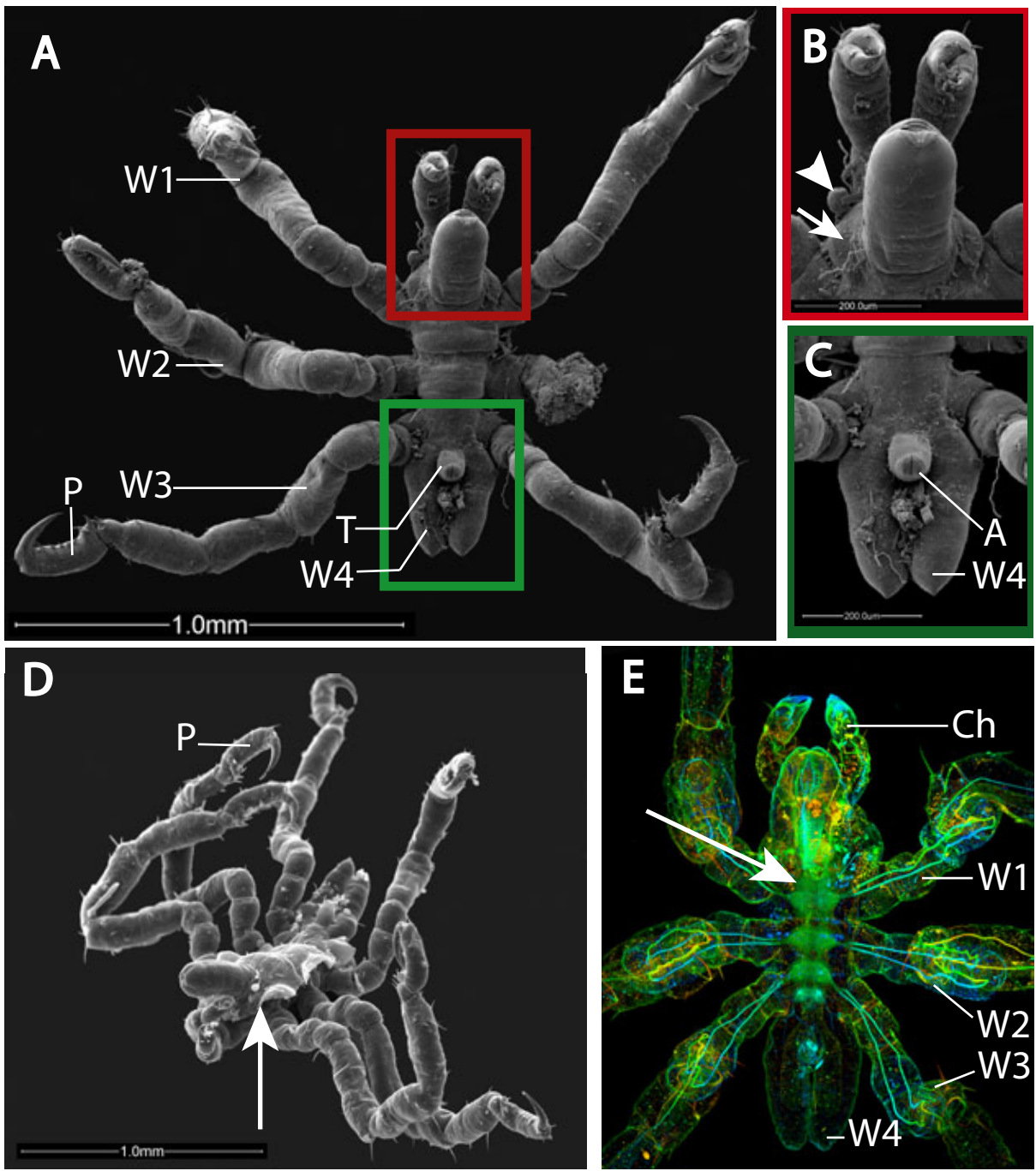

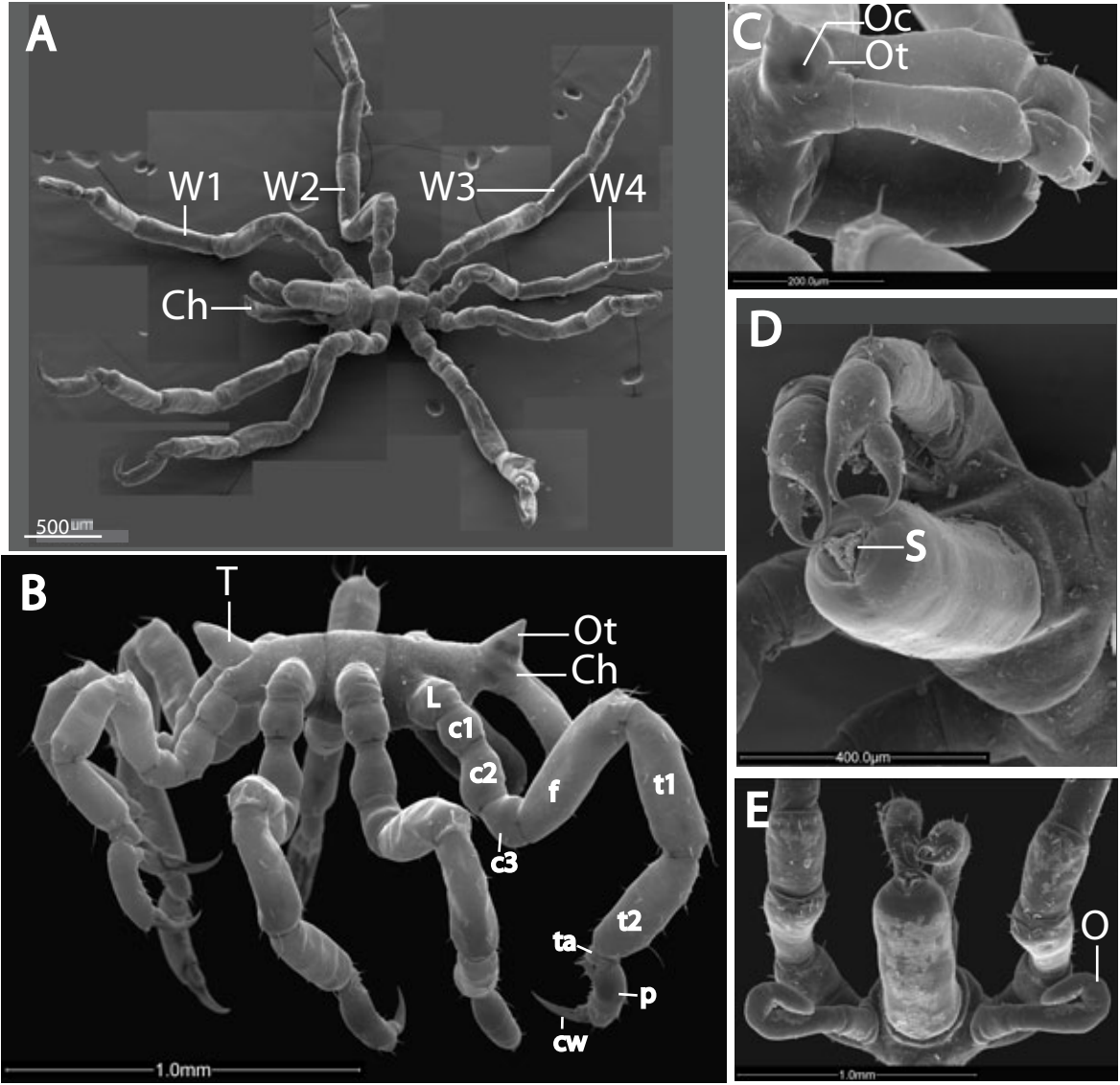

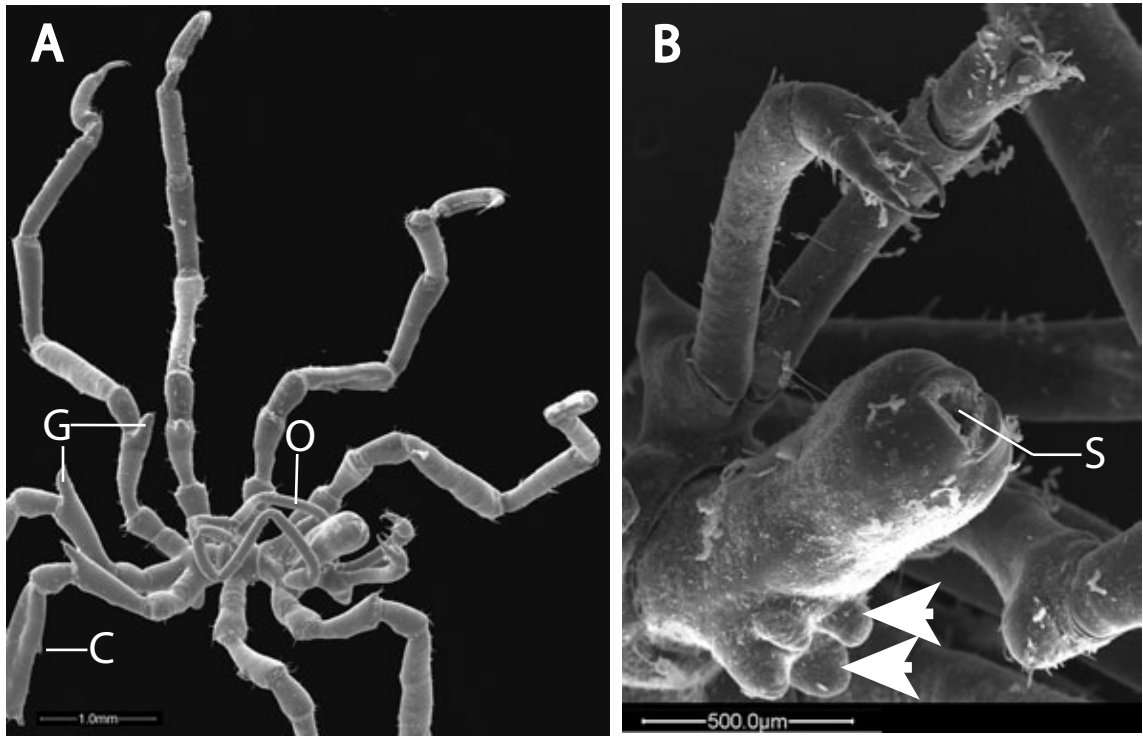